\documentclass[12pt,a4paper,twoside]{article}
  \usepackage{amsmath}
  \usepackage{amsfonts}
  \usepackage{oldgerm}
    \advance\textheight by \topskip
    \topmargin \paperheight \advance\topmargin by -\textheight 
    \divide\topmargin by 2 \advance\topmargin by -1in 
    \advance\topmargin by -\baselineskip
    \advance\topmargin by -\headsep
    \oddsidemargin \paperwidth \advance\oddsidemargin by -\textwidth
    \divide\oddsidemargin by 2 \advance\oddsidemargin by -1in
    \evensidemargin \oddsidemargin
  \def\empty{}
  \catcode`\@=11
  \@addtoreset{equation}{section}
  
  \def\sectionname{}
  \def\appendixname{Appendix}
  \def\appendix{\par
    \setcounter{section}{0}
    \setcounter{subsection}{0}
    \def\thesection{\Alph{section}}
    \let\sectionname\appendixname
  }
  \def\Section#1{
    \let\oldthesection\thesection
    \ifx\sectionname\empty\relax%
    \else\gdef\thesection{\sectionname~\oldthesection}%
    \fi%
    \section{\protect\raggedright #1}%
    \let\thesection\oldthesection%
    \advance\c@section by -1
    \refstepcounter{section}
  }
\long\def\firstofone#1{#1}
  \newcommand{\Cite}[2][]{%
    \@for\argument:=#2\do{%
      \edef\argument{\expandafter\firstofone\argument}%
      \index{\protect\cite{\argument}}%
    }%
    \def\test{#1}\ifx\test\empty\cite{#2}\else\cite[#1]{#2}\fi%
  }
  \catcode`\@=12
  \def\xref#1{(\ref{#1})}
  \newcounter{structure}[section]
  \def\thestructure{\thesection.\arabic{structure}}
  \newcounter{substructure}[structure]

  \def\prestructitem{\refstepcounter{substructure}\item}
  \def\structindent{%
    \def\finishii{\end{list}}%
    \begin{list}{}{\leftmargin1.5em \rightmargin0pt \topsep0pt}\item%
  }
  \newenvironment{structure}[1]%
  {%
    \let\finishi=\relax%
    \let\finishii=\relax%
    \def\structitem{%
      \def\finishi{\end{list}}%
      \begin{list}{{\it\roman{substructure}}\/)}{%
      \leftmargin24pt \labelsep8pt \labelwidth16pt \itemindent0pt%
      \rightmargin0pt \topsep0pt}%
      \let\structitem=\prestructitem%
      \structitem%
    }%
    \refstepcounter{structure}%
    \subsubsection*{#1 \protect\thestructure}%
  }{%
    \finishi\finishii\vskip\baselineskip%
  }
  \makeindex 
  \def\nz{{\mathbb N}}
  \def\qz{{\mathbb Q}}
  \def\rz{{\mathbb R}}
  \def\cz{{\mathbb C}}
  \def\P{{\mathbb P}}   \def\E{{\mathbb E}}
  \def\defeq{{\mathchoice{%
    \mathrel{\mskip\thickmuskip\raise.4pt\hbox{$\mathord{\displaystyle:}$}%
    \hbox{$\mathord{\displaystyle=}$}\mskip\thickmuskip}}{%
    \mathrel{\mskip\thickmuskip\raise.4pt\hbox{$\mathord{\displaystyle:}$}%
    \hbox{$\mathord{\displaystyle=}$}\mskip\thickmuskip}}{%
    \mathrel{\mskip.25\thinmuskip\raise.25pt\hbox{$\mathord{\scriptstyle:}$}%
    \hbox{$\mathord{\scriptstyle=}$}\mskip.25\thinmuskip}}{%
    \mathrel{\mskip.1\thinmuskip\raise.1pt%
             \hbox{$\mathord{\scriptscriptstyle:}$}%
    \hbox{$\mathord{\scriptscriptstyle=}$}\mskip.1\thinmuskip}%
  }}}
  \def\eqdef{{\mathchoice{%
    \mathrel{\mskip\thickmuskip\hbox{$\mathord{\displaystyle=}$}%
    \raise.4pt\hbox{$\mathord{\displaystyle:}$}\mskip\thickmuskip}}{%
    \mathrel{\mskip\thickmuskip\hbox{$\mathord{\displaystyle=}$}%
    \raise.4pt\hbox{$\mathord{\displaystyle:}$}\mskip\thickmuskip}}{%
    \mathrel{\mskip.25\thinmuskip\hbox{$\mathord{\scriptstyle=}$}%
    \raise.25pt\hbox{$\mathord{\scriptstyle:}$}\mskip.25\thinmuskip}}{%
    \mathrel{\mskip.1\thinmuskip\hbox{$\mathord{\scriptscriptstyle=}$}%
    \raise.1pt\hbox{$\mathord{\scriptscriptstyle:}$}\mskip.1\thinmuskip}%
  }}}
  \def\halmos{{%
    \hspace*{\fill}\hbox to 18pt {\hfill\vrule width 9pt height 9pt depth 0pt}%
  }\par\ignorespaces}
  \def\e{{\rm e}}
  \def\d{{\rm d}}
  \def\esssup{\mathop{\hbox{ess$\,$sup}}}
  \def\L{{\rm L}}
  \def\C{{\cal C}}
  \def\Q{{\cal Q}}
  \let\Haccent\H 
  \def\K{{\cal K}}
  \def\H{{\cal H}}
   \let\rho=\varrho \let\theta=\vartheta
  \let\Ds=\displaystyle
\begin{document}
  \pagestyle{empty} 
\vspace*{10mm}
\begin{raggedright}
{\Large\bf
   CONTINUITY PROPERTIES OF SCHR\"ODINGER\\[7pt] 
   SEMIGROUPS WITH MAGNETIC FIELDS
}\end{raggedright}
\vskip14pt \hrule width \textwidth height 1pt \vskip14pt
\begin{raggedleft}
{\large KURT BRODERIX}\\[3.5pt] 
{\it  Institut f\"ur Theoretische Physik\\
      Universit\"at G\"ottingen\\
      Bunsenstr. 9, D-37073 G\"ottingen, Germany\\
      Email: 
}     broderix@theorie.physik.uni-goettingen.de\\
\vskip14pt
{\large DIRK HUNDERTMARK\footnote{New address: 
      \parbox[t]{110mm}{\textit{%
      Department of Mathematics 253-37, California Institute of Technology\\ 
      Pasadena, CA 91125, USA. Email:} dirkh@caltech.edu}}}\\[3.5pt] 
{\it  Department of Physics, Jadwin Hall\\ 
      Princeton University\\
      P.O.Box 708, Princeton, New Jersey 08544, USA\\
      Email:
}     hdirk@princeton.edu\\
\vskip14pt
{\large HAJO LESCHKE}\\[3.5pt] 
{\it  Institut f\"ur Theoretische Physik\\
      Universit\"at Erlangen-N\"urnberg\\
      Staudtstr. 7, D-91058 Erlangen, Germany\\
      Email:
}     leschke@theorie1.physik.uni-erlangen.de\\
\vskip14pt
{Final Version}\\
\vskip3.5pt
{Mathematical~Physics~Preprint~Archive: math-ph/9808004}\\
{Published in slightly different form in 
Rev.~Math.~Phys.~{\bf 12}, 181--225 (2000)}\\
\end{raggedleft}
\vskip10mm

\noindent
The objects of the present study are one-parameter semigroups generated by
Schr\"o\-dinger operators with fairly general electromagnetic potentials. More
precisely, we allow scalar potentials from the Kato class and impose on the
vector potentials only local Kato-like conditions.  The configuration space is
supposed to be an arbitrary open subset of multi-dimensional Euclidean space;
in case that it is a proper subset, the Schr\"odinger operator is rendered
symmetric by imposing Dirichlet boundary conditions.  We discuss the continuity
of the image functions of the semigroup and show local-norm-continuity of the
semigroup in the potentials. Finally, we prove that the semigroup has a
continuous integral kernel given by a Brownian-bridge expectation. Altogether,
the article is meant to extend some of the results in B.~Simon's landmark paper
[Bull.  Amer.  Math. Soc. (N.S.) {\bf 7}, 447--526 (1982)] to non-zero vector
potentials and more general configuration spaces.
     \newpage
     \pagestyle{myheadings} 
     \markboth{K.~BRODERIX, D.~HUNDERTMARK and H.~LESCHKE}{%
               CONTINUITY PROPERTIES OF SCHR\"ODINGER SEMIGROUPS}
\section*{Contents}
\begin{description}
\item[\ref{1} ] 
  Introduction\hfill 
  \pageref{1}
\item[\ref{2} ] 
  Basic definitions and representations\hfill
  \pageref{2}
\item[\ref{3} ] 
  Continuity of the semigroup in its parameter\hfill
  \pageref{3}
\item[\ref{4} ] 
  Continuity of the image functions of the semigroup\hfill
  \pageref{4}
\item[\ref{5} ]  
  Continuity of the semigroup in the potentials\hfill
  \pageref{5}
\item[\ref{6} ] 
  Continuity of the integral kernel of the semigroup\hfill
  \pageref{6}
\item[\phantom{Appendix W}\llap{Appendix~\ref{A} }] 
  Approximability of Kato-type potentials by
  smooth functions\hfill
  \pageref{A}
\item[\phantom{Appendix W}\llap{\ref{B} }] 
  Proof of the Feynman-Kac-It\^o formula\hfill
  \pageref{B}
\item[\phantom{Appendix W}\llap{\ref{C} }] 
  Brownian-motion estimates\hfill
  \pageref{C}
\item[Acknowledgement]\hfill\pageref{NoPolishAcknowledgement}
\item[References] \hfill\pageref{NoPolishReferences}
\item[Citation index] \hfill\pageref{NoPolishCitationIndex}
\end{description}
\Section{Introduction}
\label{1}
In non-relativistic quantum physics Schr\"odinger operators with magnetic
fields play an important r\^ole \Cite{GaPa90, GaPa91}. In suitable physical
units they are given by differential expressions of the form
\begin{equation}\label{1.1}
   H_\Lambda(A,V) =
   \frac 12
   \left( -i\nabla - A(x)
   \right)^2
   + V(x)
\end{equation}
over the configuration space $\Lambda\subseteq\rz^d$. Here the scalar potential
$V$ is a real-valued function on $\Lambda$ representing the potential energy.
The vector potential $A$ is an $\rz^d$-valued function on $\Lambda$ giving rise
to the magnetic field $\nabla\times A$.  For the purposes of quantum physics
$H_\Lambda(A,V)$ has to be given a precise meaning as a self-adjoint operator
acting on the Hilbert space $\L^2(\Lambda)$ of complex-valued wave-functions on
$\Lambda$. Clearly, if $\Lambda\neq\rz^d$ this requires to impose boundary
conditions on the wave functions in the domain (of definition) of
$H_\Lambda(A,V)$.

As impressively demonstrated in Simon's landmark paper \Cite{Sim82}, many
spectral properties of $H_\Lambda(A,V)$ can efficiently be derived by studying
the Schr\"odinger semigroup $\left\{\e^{-tH_\Lambda(A,V)}\right\}_{t\ge 0}$.
The latter can be done by probabilistic techniques via the Feynman-Kac-It\^o
path-integral formula. In fact, the works \Cite{Car79,Sim82} make essential use
of this approach for the case $A=0$ and $\Lambda=\rz^d$, that is, when both the
magnetic field vanishes and the configuration space is the whole Euclidean
space.  For $A=0$ and $\Lambda\subseteq\rz^d$ some results may be found in
\Cite[Chapter~1]{Szn98}.  Another recent monograph further elucidating the
relation between ``quantum potential theory'' and Feynman-Kac processes is
\Cite{ChZh95}.

The main goal of the present work is to extend some of the key results in
\Cite{Car79,Sim82} on continuity properties to rather general $A$ and
$\Lambda$. In doing so, we follow \Cite{Sim82} in restricting ourselves to Kato
decomposable \Cite{Kat72} scalar potentials $V$.  The class of vector
potentials $A$ considered will only be restricted by local Kato-like
conditions. The configuration space will either be $\Lambda=\rz^d$ or an
arbitrary non-empty open subset $\Lambda\subset\rz^d$.  Since vector potentials
in one spatial dimension are of no physical interest, we will only consider
dimensions $d\ge 2$. As for the necessary boundary conditions for
$\Lambda\neq\rz^d$, we are only able to handle Dirichlet conditions. Apart from
that, in our opinion, the resulting class of Schr\"odinger operators with
magnetic fields is general enough to cover most systems of physical interest
with finite ground-state energy. Continuity properties of the generated
semigroups have turned out to be valuable in obtaining interesting results in
that field of mathematical physics where magnetic fields play a major r\^ole.
We mention so-called magnetic Lieb-Thirring inequalities \Cite{Erd95},
heat-kernel estimates \Cite{Erd94,Erd97a} and the existence and Lifshits
tailing of the integrated density of states of random Schr\"odinger operators
\Cite{BrHu93,Uek94a,BrHu95,Erd97b}.  In proving the statements of the present
paper we closely follow \Cite{Car79,Sim82,Szn98} using the probabilistic approach.
Since we aim at a reasonably self-contained presentation which is intelligible
for many readers, our arguments are perhaps more detailed than usual.

The reader should note that Simon's work \Cite{Sim82} is followed by several
other interesting developments in semigroup theory relating to Schr\"odinger
operators. While some of them rely on probabilistic techniques, others do not.
In a setting closest to that of the present paper one of us \Cite{Hun96,Hun98}
has considerably relaxed the conditions imposed on the vector potential $A$.
Related results with $A\neq 0$ but with weaker assumptions on the scalar
potential $V$ have been obtained in \Cite{LiMa97} building on the works
\Cite{Voi86,KoSe89} for $A=0$. Basically, in \Cite{Voi86,KoSe89,LiMa97} the
negative part of $V$ no longer needs to be infinitesimally form-bounded
relative to the unperturbed Schr\"odinger operator $H_\Lambda(0,0)$. There has
also been a lot of activity \Cite{DeCa89,DeCa94,DeCa98,LiSe96} in studying
perturbations of positivity preserving semigroups with generators more general
than $H_\Lambda(0,0)$.  Extensions into other directions investigate
perturbations by measures instead of functions
\Cite{BlMa90,AlBl91,GlRa93,GlRa94,Stu94,StVo96}. Last but not least we mention
the progress \Cite{Stu93} for Schr\"odinger-like semigroups with an underlying
configuration space which is only locally Euclidean, see also
\Cite{DeCa94,DeCa98}.

For a wealth of information on regularity and spectral properties of
Schr\"odinger operators (for $\Lambda=\rz^d$) with or without magnetic field we
recommend the recent works \Cite{Hin92,HiSt92,MoRa94}, see also \Cite{MaUe96}
and references therein. In contrast to the present paper these works do not
rely on semigroups.
\Section{Basic definitions and representations}
\label{2}
In this section we fix our basic notation and give a short compilation of the
classes of scalar potentials $V$ and vector potentials $A$ which we will
consider. Moreover, we will give a precise definition of the Schr\"odinger
operator \xref{1.1} and recall the appropriate Feynman-Kac-It\^o formula for
its semigroup.

The open ball of radius $\rho>0$ centered about the origin in the
$\nu$-dimensional Euclidean space $\rz^\nu$, $\nu\in\nz$, is denoted by
\begin{equation}\label{2.1}
   B_\rho\defeq\left\{ x\in\rz^\nu :\: |x| < \rho \right\}.
\end{equation}
Here $|x|\defeq (x \cdot x)^{1/2}$ is the norm of
$x=(x_1,\dots,x_\nu)\in\rz^\nu$ derived from the scalar product 
${x\cdot y}\defeq$ $\sum_{j=1}^\nu x_j y_j$.  Given a subset $\Omega$ of 
$\rz^\nu$ we write
\begin{equation}\label{2.2}
  \omega\mapsto\chi_{_{\scriptstyle \Omega}}(\omega)\defeq
  \begin{cases}
         1 & \text{for $\omega\in\Omega$} \\
         0 & \text{otherwise}
  \end{cases}
\end{equation}
for its indicator function.  The closure of $\Omega$ is written as
$\overline{\Omega}$ and its boundary is denoted by $\partial\Omega$.

We denote the nabla operator $\left(\frac\partial{\partial x_1}, \dots,
  \frac\partial{\partial x_\nu}\right)$ in $\rz^\nu$ as $\nabla$ and the
Lebesgue measure on the Borel subsets of $\rz^\nu$ as $\d x$. All real-valued
or complex-valued functions on $\rz^\nu$ are assumed to be Borel measurable. If
not stated otherwise, we identify functions which differ only on sets of
Lebesgue measure zero.

The Banach space $\L^p(\Omega)$ of $p^{\text{th}}$-power Lebesgue-integrable
functions ($1 \le p\le\infty$) on a Borel set $\Omega\subseteq\rz^\nu$ of
non-zero Lebesgue measure consists of complex-valued functions
$\psi:\Omega\to\cz$ such that the norm
\begin{equation}\label{2.3}
   \|\psi\|_p \defeq
   \begin{cases}\Ds
      \left(\;
         \int_\Omega |\psi(x)|^p\,\d x
      \right)^{1/p}
      &\Ds \text{\phantom{for }}\; p<\infty
   \\[-7pt] &  \text{for} \\[-7pt] \Ds
      \esssup_{x\in\Omega} |\psi(x)|
      &\Ds \text{\phantom{for }}\; p=\infty
   \end{cases}
\end{equation}
is finite. We recall that $\L^2(\Omega)$ becomes a Hilbert space when equipped
with the scalar product
\begin{equation}\label{2.4}
  \langle\phi\vert\psi\rangle \defeq
  \int_{\Omega} \left(\phi(x)\right)^{\ast}\psi(x)\,\d x .
\end{equation}
Here the star denotes complex conjugation.  The norm of an operator
$X:\L^p(\Omega)\to\L^q(\Omega)$, $1\le p,q\le\infty$ is defined as
\begin{equation}\label{2.5}
   \left\|X\right\|_{p,q}\defeq
   \sup_{\|\psi\|_p=1}\|X\psi\|_{q} .
\end{equation}
The space $\L^p_{\text{unif, loc}}(\Omega)$ of uniformly locally
$p^{\text{th}}$-power integrable functions ($1 \le p\le\infty$) consists of
functions $\psi:\Omega\to\cz$ such that the norm
\begin{equation}\label{2.6}
   \|\psi\|_{\L^p_{\text{unif, loc}}(\Omega)}\defeq
   \begin{cases}\Ds
      \sup_{x\in\rz^\nu}
      \left(\;
         \int_\Omega
            |\psi(y)|^p\,
            \chi_{_{\scriptstyle B_1}}\!(x-y)\,
         \d y
      \right)^{1/p}
      &\Ds \text{\phantom{for }}\; p<\infty
   \\[-7pt] &  \text{for} \\[-7pt] \Ds
      \Ds\esssup_{x\in\Omega} |\psi(x)|
      &\Ds \text{\phantom{for }}\; p=\infty
   \end{cases}
\end{equation}
is finite. The space $\L^p_{\text{loc}}(\Omega)$ of locally
$p^{\text{th}}$-power integrable functions ($1 \le p\le\infty$) consists of
functions $\psi:\Omega\to\cz$ such that 
$\psi\chi_{_{\scriptstyle K}}\in\L^p(\Omega)$ for all compact 
$K\subseteq\Omega$.

The positive part $f^+$ and the negative part $f^-$ of a real-valued function
$f$ on $\Omega$ are defined by
\begin{equation}\label{2.7}
   f^{\pm}(x)\defeq \sup\{{\pm}f(x),0\},\quad x\in\Omega .
\end{equation}

In case that $\Omega\subseteq\rz^\nu$ is open, we write $\C(\Omega)$ for the
space of complex-valued continuous functions on $\Omega$. The subspace of
arbitrarily often differentiable functions with compact support inside $\Omega$
is written as $\C_0^\infty(\Omega)$.

In the following $\Lambda$ denotes a fixed, non-empty, open, not necessarily
proper subset of $\rz^d$, $d\ge 2$. We stress that we will always assume $d\ge
2$. In physical terms, $\Lambda$ serves as the {\bf configuration space}. Any
complex-valued function $f$ defined {\it a priori} on $\Lambda$ will be
understood to be trivially extended to $\rz^d$ by defining $f(x)\defeq 0$ for
$x\not\in\Lambda$ without further notice.  We use this {\bf extension
  convention} to injectively embed $\L^p(\Lambda)$ into $\L^p(\rz^d)$, $1\le
p\le \infty$, and thus have $\L^p(\Lambda)\subseteq\L^p(\rz^d)$ and similarly
$\C_0^\infty(\Lambda)\subseteq\C_0^\infty(\rz^d)$ etc.  We write
\begin{equation}\label{2.8}
   g_\rho(x)\defeq
   \chi_{_{\scriptstyle B_\rho}}(x)\;
   \begin{cases}\Ds
      -\ln|x| 
      &\Ds \text{\phantom{for }}\; d=2
   \\[-7pt] & \text{for} \\[-7pt] \Ds
      |x|^{2-d} 
      &\Ds \text{\phantom{for }}\; d\ge 3
   \end{cases}
\end{equation}
for the repulsive Newton-Coulomb potential on $\rz^d$ truncated outside the
Ball $B_\rho$.
\begin{structure}{Definition}\label{2-1}\it\structindent
  A {\bf scalar potential} is a real-valued function $V$ on $\Lambda$.  The
  function $V$ is said to be in the {\bf Kato class} $\K(\rz^d)$ if
  \begin{equation}\label{2.9}
    \lim_{\rho\downarrow 0}\sup_{x\in\rz^d}\:
    \int g_\rho(x-y)\,|V(y)|\,\d y
    = 0.
  \end{equation}
  The function $V$ is said to be in the {\bf local Kato class}
  $\K_{\text{loc}}(\rz^d)$ if $V\chi_{_{\scriptstyle K}}\in\K(\rz^d)$ for all
  compact $K\subset\rz^d$. The function $V$ is said to be {\bf Kato
    decomposable}, in symbols $V\in\K_{\pm}(\rz^d)$, if
  $V^+\in\K_{\text{loc}}(\rz^d)$ and $V^-\in\K(\rz^d)$.  The {\bf Kato norm}
  is given by
  \begin{equation}\label{2.10}
    \|V\|_{\K(\rz^d)}\defeq
    \sup_{x\in\rz^d}\:
    \int g_1(x-y)\:|V(y)|\,\d y.
  \end{equation}
\end{structure}
Note that due to the above extension convention it is reasonable to state
$V\in\K(\rz^d)$ even if $V$ is {\it a priori} only defined on
$\Lambda\subseteq\rz^d$.

The following remarks are borrowed from \Cite[Chapter~1.2]{CyFr87},
\Cite[Section~4]{AiSi82} and \Cite[Chapter~3]{ChZh95}.
\begin{structure}{Remarks}\label{2-2}\structindent
\structitem\label{2-2-i} To find out whether or not a function belongs to
  the Kato class or local Kato class the following inclusions may be helpful.
  For $p>\frac{d}{2}$ one has
  \begin{gather}\label{2.11}
    \L^p_{\text{unif, loc}}(\rz^d)
    \subset \K(\rz^d) \subset
    \L^1_{\text{unif, loc}}(\rz^d),
  \\ \label{2.12}
    \L^p_{\text{loc}}(\rz^d)
    \subset \K_{\text{loc}}(\rz^d) \subset
    \L^1_{\text{loc}}(\rz^d) .
  \end{gather}  
\structitem\label{2-2-ii} The Kato class is complete with respect to the
  Kato norm \Cite[Erratum]{Sim82}, \Cite[Section~5]{Voi86}, that is,
  $\K(\rz^d)$ is a Banach space. The inclusions \xref{2.11} hold also with
  respect to convergence in norm, that is, convergence in
  \hbox{$\|\bullet\|_{\L^p_{\text{unif, loc}}(\rz^d)}$} implies convergence in
  \hbox{$\|\bullet\|_{\K(\rz^d)}$} which in turn implies convergence in
  \hbox{$\|\bullet\|_{\L^1_{\text{unif, loc}}(\rz^d)}$}, where
  $p>\frac{d}{2}$.  \structitem\label{2-2-iii} Let $2H_\Lambda(0,0)$ denote
  the self-adjoint Friedrichs-extension on the Hilbert space $\L^2(\Lambda)$
  of the negative Laplacian $-{\nabla\cdot\nabla}$ on $\C^\infty_0(\Lambda)$,
  that is, $-2H_\Lambda(0,0)$ is the usual Laplacian on $\Lambda$ with
  Dirichlet boundary conditions. Then any $V\in\K(\rz^d)$ is infinitesimally
  form-bounded \Cite[Definition~p.~168]{ReSi75} relative to $H_\Lambda(0,0)$
  \Cite[Corollary to Theorem 3.25]{ChZh95}. Alternatively, this follows from
  \Cite[Proposition~2.1]{Kat72} or \Cite[Theorem~4.7]{AiSi82} used with the
  fact that, by the extension convention, the form domain of $H_\Lambda(0,0)$
  is a subset of the form domain of $H_{\rz^d}(0,0)$.  The class $\K(\rz^d)$
  is nearly maximal with respect to infinitesimal form-boundedness
  \Cite[Theorem~1.12]{CyFr87}.  Consequently, the set of Kato-decomposable
  functions should contain all physically relevant scalar potentials, which
  lead to Schr\"odinger operators \xref{1.1} with a finite ground-state
  energy.  \structitem\label{2-2-iv} An example illustrating the admissible
  local singularities of scalar potentials in $\K(\rz^d)$ for $d\ge 3$ is
  \begin{equation}\label{2.13}
    V_\mu(x) \defeq
    \frac{\Theta_{1/2}(x)
         }{|x|^2\,\left|\ln|x|\right|^\mu
         }
   ,\quad \mu>0,
  \end{equation}
  where $\Theta_{1/2}$ is some real-valued function in $\C^\infty_0(\rz^d)$
  which vanishes outside the ball $B_{3/4}$ and furthermore obeys
  $\Theta_{1/2}(x)=1$ for $|x|<1/2$.  According to \Cite[Example (a)
  p.8]{CyFr87}, \Cite[Proposition~4.10]{AiSi82} one has the equivalence
  \begin{equation}\label{2.14}
    V_\mu\in\K(\rz^d)\Leftrightarrow\mu>1.
  \end{equation}
  Note that, for $d=3$, $V_\mu$ obeys the Rollnik condition
  \Cite[Chapter~I]{Sim71} whenever $\mu>\frac 12$, see
  \Cite[Example~I.6.3]{Sim71}. Finally, $V_\mu$ is infinitesimally form-bounded
  relative to $H_{\rz^d}(0,0)$ for all $\mu>0$.
\end{structure}
Useful for handling Kato-decomposable functions is the following possibility
to approximate them by nice functions.
\begin{structure}{Proposition}\label{2-3}\structindent\it
  Let $V\in\K_{\pm}(\rz^d)$. Then there is a sequence
  $\left\{V_n\right\}_{n\in\nz}\subset\C^\infty_0(\rz^d)$ such that
  \begin{equation}\label{2.15}
    \lim_{n\to\infty}
    \left\|(V-V_n)\:\chi_{_{\scriptstyle K}}\right\|_{\K(\rz^d)}
    = 0
  \end{equation}
  for all compact $K\subset\rz^d$ and
  \begin{equation}\label{2.16}
    \sup_{n\in\nz}\sup_{x\in\rz^d}\:
    \int g_\rho(x-y)\:V^-_n(y)\,\d y
    \le
    \sup_{x\in\rz^d}\:
    \int g_\rho(x-y)\:V^-(y)\,\d y
  \end{equation}
  for all $0<\rho\le 1$.
\end{structure}
This approximability and the idea of its proof have been stated in
\Cite[\S~B.10]{Sim82}. The approximating sequence $\{V_n\}$ may be constructed
from $V$ in the following standard way. The potential $V$ is smoothly truncated
outside an increasingly large ball and then mollified by convolution with an
approximate delta function in $\C^\infty_0(\rz^d)$. Since $V\in\K_{\pm}(\rz^d)$
is locally integrable, the resulting approximations are in
$\C^\infty_0(\rz^d)$. Then patiently estimating shows that \xref{2.15} and
\xref{2.16} can be fulfilled. The details can be found in Appendix~\ref{A}.

We now define the classes of vector potentials to be dealt with in the sequel.
\begin{structure}{Definition}\label{2-4}\structindent\it
  A {\bf vector potential} is an $\rz^d$-valued function $A$ on $\Lambda$. A
  vector potential $A$ is said to be in the class $\H(\rz^d)$, if its squared
  norm $A^2\defeq A\cdot A$ and its divergence $\nabla\cdot A$, considered as a
  distribution on $\C^\infty_0(\rz^d)$, are both in $\K(\rz^d)$. It is said to
  be in the class $\H_{\text{loc}}(\rz^d)$, if both $A^2$ and $\nabla\cdot A$
  are in $\K_{\text{loc}}(\rz^d)$.
\end{structure}
\unskip
\begin{structure}{Remarks}\label{2-5}\structindent
\structitem\label{2-5-i} As it should be from the physical point of view,
  only local regularity conditions are required for a vector potential to lie
  in $\H_{\text{loc}}(\rz^d)$. Moreover, since once continuously differentiable
  vector potentials are included, in symbols
  $\left(\C^1(\rz^d)\right)^d\subset\H_{\text{loc}}(\rz^d)$, most physically
  relevant vector potentials are covered. From this point of view the class
  $\H(\rz^d)$ is of less interest, because, for example, any vector potential
  $A\in\left(\C^1(\rz^3)\right)^3$ giving rise to a constant magnetic field
  $\nabla\times A$ lies in $\H_{\text{loc}}(\rz^3)$ but not in $\H(\rz^3)$.
  Therefore, we try to avoid global regularity assumptions as far as possible,
  that is, we aspire after results valid for vector potentials in
  $\H_{\text{loc}}(\rz^d)$.  \structitem\label{2-5-ii} With regard to the
  one-parameter family of vector potentials considered in
  \Cite[Theorem~6.2]{CyFr87}, we note that the spectrum of the associated
  Schr\"odinger operator dramatically changes its character precisely at that
  parameter value where the border between $\H(\rz^d)$ and
  $\H_{\text{loc}}(\rz^d)$ is reached.  \structitem\label{2-5-iii} The local
  singularities admissible for vector potentials in $\H(\rz^d)$, $d\ge 3$, are
  illustrated by the example
  \begin{equation}\label{2.17}
    A_\mu(x) \defeq
    \frac{x}{|x|}
    \left(V_\mu(x)\right)^{1/2},
  \end{equation}
  where $V_\mu$ is defined in \xref{2.13}. According to Remark~\ref{2-2-iv}
  one has ${A_\mu^2\in\K(\rz^d)}\Leftrightarrow {\mu>1}$, but
  \begin{equation}\label{2.18}
    A_\mu\in\H(\rz^d)\Leftrightarrow\mu>2,
  \end{equation}
  as can be seen by explicitly calculating $\nabla\cdot A_\mu$.
\end{structure}
Similar to Kato-decomposable scalar potentials, vector potentials in
$\H_{\text{loc}}(\rz^d)$ can be approximated by smooth functions.
\begin{structure}{Proposition}\label{2-6}\structindent\it
  Let $A\in\H_{\text{loc}}(\rz^d)$. Then there is a sequence
  $\{A_m\}_{m\in\nz}\subset\left(\C^\infty_0(\rz^d)\right)^d$ such that
  \begin{equation}\label{2.19}
    \lim_{m\to\infty}
    \left\|
      (A-A_m)^2\:\chi_{_{\scriptstyle K}}
    \right\|_{\K(\rz^d)} = 0
  \end{equation}
  and
  \begin{equation}\label{2.20}
    \lim_{m\to\infty}
    \left\|
      (\nabla\cdot A-\nabla\cdot A_m)\:
      \chi_{_{\scriptstyle K}}
    \right\|_{\K(\rz^d)} = 0
  \end{equation}
  for all compact $K\subset\rz^d$.
\end{structure}
The proof is similar to the one of Proposition~\ref{2-3} and is given in
Appendix~\ref{A}.

The next topic is the precise construction of the {\bf Schr\"odinger operator}
\xref{1.1} as a self-adjoint operator acting on the Hilbert space
$\L^2(\Lambda)$. Here we use the definition via forms
\Cite[Section~VIII.6]{ReSi80}, \Cite[Section~X.3]{ReSi75},
\Cite[Section~1.1]{CyFr87}.

In a first step we consider $A\in\H_{\text{loc}}(\rz^d)$ and
$V^+\in\K_{\text{loc}}(\rz^d)$.  The sesquilinear form
\begin{equation}\label{2.21}
  h_\Lambda^{A,V^+}:
  \C^\infty_0(\Lambda) \times \C^\infty_0(\Lambda) \to \cz
  \, ,\quad
  (\phi,\psi) \mapsto   h_\Lambda^{A,V^+}(\phi,\psi)
\end{equation}
where
\begin{equation}\label{2.22}\!\!\!\!\!\!\!\!\!\!\!\!\!\!\!\!\!\!\!\!\!\!\!\!
  h_\Lambda^{A,V^+}(\phi,\psi) \defeq
  \langle
    (V^+)^{\frac{1}{2}}\phi
  \,\vert\,
    (V^+)^{\frac{1}{2}}\psi
  \rangle
  +
  \frac{1}{2}\sum_{j=1}^d
  \langle
    (-i\nabla-A)_j\phi
  \,\vert\,
    (-i\nabla-A)_j\psi
  \rangle           \!\!\!\!\!\!\!\!
\end{equation}
is densely defined in $\L^2(\Lambda)$ and non-negative. For $\Lambda=\rz^d$ the
closure of the form \xref{2.21} has form domain
\begin{equation}\label{2.23}\!\!\!\!\!\!\!\!\!\!\!\!\!\!\!\!\!\!
  \Q\!\left(h_{\rz^d}^{A,V^+}\right) :=
  \left\{
    \psi\in\L^2(\rz^d):\:
    (-i\nabla-A)\psi\in\left(\L^2(\rz^d)\right)^d,\,
    (V^+)^{\frac{1}{2}}\psi\in\L^2(\rz^d)
  \right\} \!\!\!\!\!\!\!\!\!
\end{equation}
see \Cite{Sim79b}, \Cite[Theorem~1.13]{CyFr87}. For general open
$\Lambda\subseteq\rz^d$ the completion $\Q\!\left(h_\Lambda^{A,V^+}\right)$ of
$\C_0^\infty(\Lambda)$ with respect to the form norm
\begin{equation}\label{2.24}
  \|\psi\|_{h_\Lambda^{A,V^+}}  \defeq
  \sqrt{\|\psi\|_2^2+h_\Lambda^{A,V^+}(\psi,\psi)}
\end{equation}
cannot exceed $\L^2(\rz^d)$ because the form is closable for $\Lambda=\rz^d$.
Since $\L^2(\Lambda)$ is a closed subspace of $\L^2(\rz^d)$ the completion is
even a subspace of $\L^2(\Lambda)$. Thus the form \xref{2.21} extended to the
domain $\Q\!\left(h_\Lambda^{A,V^+}\right)$ is non-negative and closed.
According to \Cite[Theorem~VIII.15]{ReSi80} it therefore defines a unique
self-adjoint operator on $\L^2(\Lambda)$ denoted as $H_\Lambda(A,V^+)$.

Before proceeding with the construction of the Schr\"odinger operator we note
that the just-defined Schr\"odinger operator $H_\Lambda(A,0)$,
$A\in\H_{\text{loc}}(\rz^d)$, obeys the so-called diamagnetic inequality
\begin{equation}\label{2.25}
  |\e^{-tH_\Lambda(A,0)}\psi|
  \le\e^{-tH_\Lambda(0,0)}|\psi|,\quad \psi\in\L^2(\Lambda),\; t\ge 0.
\end{equation}
For $\Lambda=\rz^d$ the inequality may be found in
\Cite[Proposition~B.13.1]{Sim82}, \Cite[Theorem~2.3]{AvHe78}.  For general open
$\Lambda\subseteq\rz^d$ the estimate \xref{2.25} is given in
\Cite[Corollary~3.6]{PeSe81} and is a special case of
\Cite[Corollary~2.3]{LiMa97}. Moreover, we will get it in the version
\xref{2.25} from Lemma~\ref{B-5} in Appendix~\ref{B} devoted to a proof of the
Feynman-Kac-It\^o formula below.

In the second step we consider $A\in\H_{\text{loc}}(\rz^d)$ and
$V\in\K_{\pm}(\rz^d)$.  According to Remark~\ref{2-2-iii} $V^-\in\K(\rz^d)$ is
infinitesimally form-bounded relative to $H_\Lambda(0,0)$.  This and the
diamagnetic inequality \xref{2.25} imply that $V^-$ is infinitesimally
form-bounded relative to $H_\Lambda(A,0)\le H_\Lambda(A,V^+)$.  In the case
$\Lambda=\rz^d$ the last statement is proven in \Cite[Theorem~2.5]{AvHe78},
\Cite[Theorem~15.10]{Sim79a}. For general open $\Lambda\subseteq\rz^d$ it is
proven in \Cite[Proposition~3.7]{PeSe81}. All three proofs are virtually
identical.

Due to the established form-boundedness the KLMN theorem
\Cite[Theorem~X.17]{ReSi75} is applicable and ensures that
\begin{equation}\label{2.26}
\begin{split}\Ds & \Ds
  h_\Lambda^{A,V}:
  \Q\!\left(h_\Lambda^{A,V^+}\right) \times 
  \Q\!\left(h_\Lambda^{A,V^+}\right)\to \cz,
\\ \Ds & \Ds
  (\phi,\psi) \mapsto h_\Lambda^{A,V}(\phi,\psi) \defeq 
  h_\Lambda^{A,V^+}(\phi,\psi) -
  \langle
    (V^-)^{\frac{1}{2}}\phi
  \,\vert\,
    (V^-)^{\frac{1}{2}}\psi
  \rangle
\end{split}
\end{equation}
is a closed sesquilinear form bounded from below and with form core
$\C_0^\infty(\Lambda)$. The associated semi-bounded self-adjoint operator is
denoted as $H_\Lambda(A,V)$.
\begin{structure}{Remarks}\label{2-7}\structindent
\structitem\label{2-7-i} Of course, the above construction of
  $H_\Lambda(A,V)$ works for $A^2,V^+\in\L^1_{\text{loc}}(\rz^d)$ and $V^-$
  form-bounded relative to $H_\Lambda(A,V^+)$ with bound strictly smaller than
  1.  If we even assume $A^2,\nabla\cdot A,V^+\in\L^2_{\text{loc}}(\rz^d)$ the
  construction of $H_\Lambda(A,V)$ is a bit easier, since then the form
  \xref{2.21} comes directly from the non-negative symmetric operator
  \begin{equation}\label{2.27}
    \C^\infty_0(\Lambda) \to \L^2(\Lambda), \quad
    \psi \mapsto \frac{1}{2} 
    \left( -i\nabla - A \right)^2\psi + V^+ \psi
  \end{equation}
  and is therefore closable \Cite[Theorem~X.23]{ReSi75}.
\structitem\label{2-7-ii} $\C^\infty_0(\rz^d)$ is not only a form core but
  even an operator core for $H_{\rz^d}(A,V)$, if in addition to
  $A\in\H_{\text{loc}}(\rz^d)$, $V\in\K_{\pm}(\rz^d)$ one has $A^2,\nabla\cdot
  A,V\in\L^2_{\text{loc}}(\rz^d)$.  This follows from
  \Cite[Theorem~2.5]{HiSt92}.  In \Cite[Theorem~B.13.4]{Sim82} the statement
  without the restriction $\nabla\cdot A\in\L^2_{\text{loc}}(\rz^d)$ is
  incorrectly ascribed to \Cite{LeSi81}.  Other criteria for the essential
  self-adjointness of $H_{\rz^d}(A,V)$ on $\C^\infty_0(\rz^d)$ may be found for
  instance in \Cite[Theorem~3]{LeSi81} and, more generally, in
  \Cite[Corollary~1.4]{Lei83}.  
\structitem\label{2-7-iii} If $\Lambda\neq\rz^d$ the above construction of
  $H_\Lambda(A,V)$ corresponds, roughly speaking, to imposing Dirichlet
  boundary conditions on $\partial\Lambda$ in order to render \xref{1.1}
  formally self-adjoint. In fact, for $A=0$ one has
  $\Q\!\left(h_\Lambda^{0,V}\right) = \Q\!\left(V^+\right) \cap
  \smash{\overset{\circ}{\text{H}}}^1(\Lambda)$ where
  $\smash{\overset{\circ}{\text{H}}}^1(\Lambda)$ is the Sobolev space of
  functions in $\L^2(\Lambda)$ vanishing on $\partial\Lambda$ in distributional
  sense and having a square-integrable distributional gradient, see, for
  example, \Cite[Section~2.3]{EgSh92}.
\end{structure}
We are going to recall the Feynman-Kac-It\^o formula for the semigroup
$\{\e^{-tH_\Lambda(A,V)}\}_{t\ge 0}$ generated by the above-defined
Schr\"odinger operator $H_\Lambda(A,V)$.

We denote by $\P_{x}$ Wiener's probability measure associated with standard
{\bf Brownian motion} $w$ on $\rz^d$ having diffusion constant $\frac 12$ and
almost surely continuous paths $s\mapsto w(s)$ starting from $x\in\rz^d$, that
is, $w(0)=x$. The induced expectation is written as $\E_{x}$.

According to \Cite[Theorem~4.5]{AiSi82} the Kato class is conveniently
characterized in terms of Brownian motion by the following equivalence
\begin{equation}\label{2.28}
  f\in\K(\rz^d)\Leftrightarrow
  \lim_{t\downarrow 0}\sup_{x\in\rz^d}
  \E_{x}\!\left[\int_0^t|f(w(s))|\,\d s\right]
  = 0 .
\end{equation}
This is a consequence of the chain of inequalities
\begin{equation}\label{2.29}
  \begin{split}\Ds & \Ds
    C_1(d) \int g_{t^{\kappa(d)}}(x-y)\,|f(y)|\,\d y 
  \\ \Ds \le\; & \Ds
    \E_{x}\!\left[\int_0^t|f(w(s))|\,\d s\right] 
  \\ \Ds \le\; & \Ds
    C_2(d) \int g_{\rho}(x-y)\,|f(y)|\,\d y
    +
    C_3(d,\rho,t)\, 
    \|f\|_{\L^1_{\text{unif, loc}}(\rz^d)} ,
  \end{split}
\end{equation}
which hold for all $d\ge 2$, $0<t<1$, $0<\rho<\frac{1}{2}$. Here we have set
$\kappa(2)\defeq 1$ and $\kappa(d)\defeq \frac 12$ for $d\ge 3$ and
$C_1(d),C_2(d)$ are strictly positive constants. Moreover, $C_3$ is a function
obeying $\lim_{t\downarrow 0} C_3(d,\rho,t) =0$ for all $d\ge 2$,
$0<\rho<\frac{1}{2}$. This chain of inequalities is implicit in the proof of
\Cite[Theorem~4.5]{AiSi82}.

Since, almost surely, Brownian-motion paths stay for finite times in bounded
regions, the equivalence \xref{2.28} proves the implication
\begin{equation}\label{2.30}
  f\in\K_{\text{loc}}(\rz^d)\Rightarrow
  \P_{x}\!\left\{\int_0^t|f(w(s))|\,\d s <\infty\right\} = 1,
  \quad x\in\rz^d,\; t\ge 0.
\end{equation}
Confer the proof of Lemma~\ref{C-8} of Appendix~\ref{C}.
\begin{structure}{Remarks}\label{2-8}\structindent
\structitem\label{2-8-i} The implication \xref{2.30} explains with the help
  of \Cite[Section~4.3]{Fri75} or \Cite[Definition~3.2.23]{KaSh91} that the
  stochastic line integral in the sense of It\^o
  \begin{equation}\label{2.31}
    t \mapsto \int_0^t A(w(s))\cdot\d w(s)
  \end{equation}
  of a vector potential $A\in\H_{\text{loc}}(\rz^d)$ is for all
  $x=w(0)\in\rz^d$ a well-defined stochastic process possessing a continuous
  version.  
\structitem\label{2-8-ii} For a vector potential $A\in\H_{\text{loc}}(\rz^d)$
  and a scalar potential $V\in\K_{\pm}(\rz^d)$ the potentials' part of the
  Euclidean {\bf action}
  \begin{equation}\label{2.32}
    t\mapsto S_t(A,V|w), \quad t\ge 0,
  \end{equation}
  where
  \begin{equation}\label{2.33}\!\!\!\!\!\!\!\!
    \begin{split}\Ds
      S_t(A,V|w) \, \defeq & \Ds \: i\int_0^t A(w(s))\cdot\d w(s)
      + \frac{i}{2} \int_0^t (\nabla\cdot A)\,(w(s))\,\d s 
    \\ &\Ds
      \hbox{}+
      \int_0^t V(w(s))\,\d s\\
    \end{split}
  \end{equation}
  is for all $x=w(0)\in\rz^d$ a well-defined complex-valued stochastic process
  possessing a continuous version.  This follows from the preceding remark and
  \xref{2.30}, confer \Cite[Lemma~3.1]{Car79}.
\end{structure}
Our basic tool will be the following variant of the {\bf Feynman-Kac-It\^o
  formula}.
\begin{structure}{Proposition}\label{2-9}\structindent\it
  Let $A\in\H_{\text{loc}}(\rz^d)$, $V\in\K_{\pm}(\rz^d)$ and
  $\Lambda\subseteq\rz^d$ open.  Then for any $\psi\in\L^2(\Lambda)$ and $t\ge
  0$ one has
  \begin{equation}\label{2.34}
    \left(\e^{-tH_\Lambda(A,V)}\psi\right)(x) =
    \E_{x}\!\left[
      \e^{-S_t(A,V|w)}\,\Xi_{\Lambda,t}(w)\,\psi(w(t))
    \right]
  \end{equation}
  for almost all $x\in\Lambda$, where
  \begin{equation}\label{2.35}
    \Xi_{\Lambda,t}(w) \defeq
    \begin{cases}
      1 & \text{if $w(s)\in\Lambda$ for all $0<s\le t$} \\
      0 & \text{otherwise.}
    \end{cases}
  \end{equation}
\end{structure}
For $\Lambda=\rz^d$, $\nabla\cdot A = 0$ the proposition is a special case of
\Cite[Theorem~15.5]{Sim79a}. The proof given there can be extended easily to
$\Lambda=\rz^d$, $\nabla\cdot A\in\L^1_{\text{loc}}(\rz^d)$. For
$\Lambda\subseteq\rz^d$, $A=0$, $V=0$ the proposition is equivalent to
\Cite[Theorem~3]{Sim78a} and \Cite[Theorem~21.1]{Sim79a}.  Finally, for
$\Lambda\subseteq\rz^d$, $A=0$, $V^+\in\K_{\text{loc}}(\rz^d)$, $V^-=0$, the
proposition follows from \Cite[Proposition~1.3.3, Theorem~1.4.11]{Szn98}.  We
have not found a proof for the above setting in the literature. Therefore we
give a detailed proof in Appendix~\ref{B}.  There we use Proposition~\ref{2-9}
for $\Lambda=\rz^d$ to get \xref{2.34} for $\Lambda\subset\rz^d$,
$A\in\H_{\text{loc}}(\rz^d)$ and $V\in\L^\infty(\rz^d)$ in a way patterned
after the proof of \Cite[Theorem~3]{Sim78a}, \Cite[Theorem~21.1]{Sim79a}.
Eventually we generalize the achieved result to the assumptions of the
proposition by an approximation technique taken from the proof of
\Cite[Theorem~6.2]{Sim79a}.

With considerably more effort it is possible to prove a Feynman-Kac-It\^o
formula for still more general vector potentials, see \Cite{Hun96,Hun98}.
There is even a Feynman-Kac-It\^o formula under virtually minimal conditions
\Cite{PeSe81}, which suffers, however, from being less explicit than
\xref{2.34}.

From the triangle inequality and the proof of \Cite[Proposition~3.1]{Car79} or
\Cite[Theorem~B.1.1]{Sim82} the right-hand side of \xref{2.34} -- considered as
a function of $x$ -- is seen to lie in $\L^q(\Lambda)$ for any
$\psi\in\L^p(\Lambda)$ where $t>0$, $1\le p\le q\le\infty$.  This statement is
also a special case of Lemma~\ref{C-1}.  Hence the right-hand side of
\xref{2.34} can be used to define an operator
$T^t_{p,q}:\L^p(\Lambda)\to\L^q(\Lambda)$.  By the same argument this operator
is bounded, that is,
\begin{equation}\label{2.36}
  \left\|T^t_{p,q}\right\|_{p,q} < \infty,
  \quad t>0,\; 1\le p\le q\le\infty.
\end{equation}
The family of operators $\{T^t_{p,q}\}_{t\ge 0,\, 1\le p\le q\le\infty}$ thus
obtained constitutes a one-parameter {\bf semigroup} in the sense that
\begin{equation}\label{2.37}
  T^{0}_{p,p} = 1, \quad
  T^{s+t}_{p,r} = T^s_{q,r}\circ T^t_{p,q},\quad
  s,t>0, \; 1\le p\le q\le r\le\infty.
\end{equation}
This follows from the Markov properties of Brownian motion and of the It\^o
integral in \xref{2.33}. Moreover, the family is {\bf self-adjoint} in the
sense that the dual mapping of $T^t_{p,q}$ is $T^t_{p',q'}$ where the indices
$p',q'$ dual to $p,q$ are defined as usual through $\frac{1}{p'}+\frac{1}{p}
\defeq 1 \eqdef \frac{1}{q'}+\frac{1}{q}$.

In the next section we will see that $\{T^t_{p,p}\}_{t\ge 0}$ is a strongly
continuous semigroup for all finite $1\le p<\infty$.  Nevertheless, for
mnemonic reasons, we will always write $\e^{-tH_\Lambda(A,V)}$ instead of
$T^t_{p,q}$ for all $1\le p\le q\le\infty$.

Within this notation the triangle inequality applied to the right-hand side of
\xref{2.34} generalizes the {\bf diamagnetic inequality} \xref{2.25} to
\begin{equation}\label{2.38}
  |\e^{-tH_\Lambda(A,V)}\psi|
  \le \e^{-tH_\Lambda(0,V)}|\psi|,\quad \psi\in\L^p(\Lambda),
\end{equation}
confer \Cite[Theorem~B.13.2]{Sim82}, \Cite[Equation~(2.13)]{LiMa97}.

For vanishing magnetic field $A=0$ the semigroup is increasing in $\Lambda$ in
the sense that for $\Lambda\subseteq\Lambda'\subseteq\rz^d$, $\Lambda$,
$\Lambda'$ open, one has
\begin{equation}\label{2.39}
  \e^{-tH_\Lambda(0,V)}\chi_{_{\scriptstyle \Lambda}} \psi \le 
  \e^{-tH_{\Lambda'}(0, V)} \psi  
\end{equation}
for all $\psi\ge 0$, $\psi\in\L^p(\Lambda')$, $1\le p\le\infty$ and $t\ge 0$.
Because $\Xi_{\Lambda,t}\le\Xi_{\Lambda',t}$, this follows from \xref{2.34} by
inspection.

The above-mentioned boundedness of $\e^{-tH_\Lambda(A,V)}$ can now be sharpened
to the statement that for given $V\in\K_{\pm}(\rz^d)$ there are real constants
$C$ and $E$, independent of $t$, $p$ and $q$, so that
\begin{equation}\label{2.40}
\begin{split}\Ds & \Ds
  \left\|\e^{-tH_\Lambda(A,V)}\right\|_{p,q} \le
  \left\|\e^{-tH_\Lambda(0,V)}\right\|_{p,q} \le
  \left\|\e^{-tH_{\rz^d}(0,V)}\right\|_{p,q} 
\\ \Ds & \Ds \le
  \begin{cases}\Ds
    C \: t^{-(q-p)d/2pq} \: \e^{-tE} &\Ds \text{\phantom{for }} 1\le
    p < q \le\infty \\[-7pt] & \text{for} \\[-7pt] \Ds C \: \e^{-tE}
    &\Ds \text{\phantom{for }} 1\le p = q \le\infty
  \end{cases}
\end{split}
\end{equation}
for all $A\in\H_{\text{loc}}(\rz^d)$. The first inequality follows from
\xref{2.38}, confer \Cite[Corollary~B.13.3]{Sim82}, the middle inequality is a
consequence of \xref{2.39} and the last inequality is given in
\Cite[Equation~(B.11)]{Sim82}. For related estimates see also
\Cite[Section~1]{AiSi82} and references therein. Furthermore, $C$ and $E$ can
be chosen \Cite[Theorem~B.5.1]{Sim82} such that $E$ is any number smaller than
the infimum of the $\L^2(\rz^d)$-spectrum of $H_{\rz^d}(0,V)$.

Eventually, we turn to the notion of a configuration space with a regular
boundary.
\begin{structure}{Definition}\label{2-10}\structindent\it
  A set $\Lambda\subset\rz^d$ is called {\bf regular}, if it is open and
  \begin{equation}\label{2.41}
    \E_{x}\!\left[
      \Xi_{\Lambda,t}(w)
    \right] =0
  \end{equation}
  for all $x\in\partial\Lambda$, $t>0$. The set $\Lambda\subset\rz^d$ is
  called {\bf uniformly regular}, if it is open and
  \begin{equation}\label{2.42}
    \lim_{\tau\downarrow 0}\, \sup_{x\in\partial\Lambda}
    \E_{x}\!\left[
      \Xi_{\Lambda,t-\tau}(w(\bullet + \tau))
    \right] =0
  \end{equation}
  for all $t>0$. Furthermore, it is understood that $\Lambda=\rz^d$ is both
  regular and uniformly regular.
\end{structure}
\unskip
\begin{structure}{Remarks}\label{2-11}\structindent
\structitem\label{2-11-i} The above definition of regularity is equivalent to
  the standard one \Cite[Definition~II.1.9]{Bas95}, \Cite[Section~2.3]{PoSt78},
  which employs first exit times. More precisely, $\Lambda\subset\rz^d$ is
  regular if and only if it is open and
  \begin{equation}\label{2.43}
    \P_{x}\!\left\{
      \inf \{ s>0:w(s)\not\in\Lambda\} = 0
    \right\} = 1
  \end{equation}
  for all $x\in\partial\Lambda$. To prove this assertion we first mention that
  by the definition \xref{2.35} of $\Xi_{\Lambda,t}(w)$, \xref{2.43} implies
  \xref{2.41}. To show the opposite direction, note that
  \begin{equation}\label{2.44}
    t \mapsto  \{ w(s)\in\Lambda\text{ for all }0<s\le t\} 
  \end{equation}
  is decreasing in the sense of set inclusion. Therefore, \xref{2.43} is
  equivalent to
  \begin{equation}\label{2.45}
    \P_x\!\left\{
      \bigcup_{t>0, t\in\qz}\{ w(s)\in\Lambda\text{ for all }0<s\le t\} 
    \right\} =0,
  \end{equation}
  where $\qz$ denotes the set of rational numbers. Due to \xref{2.35},
  \xref{2.41} implies \xref{2.45}.  
\structitem\label{2-11-ii} Using dominated convergence, one checks that
  \xref{2.41} is equivalent to
  \begin{equation}\label{2.46}
    \lim_{\tau\downarrow 0}
    \E_{x}\!\left[
      \Xi_{\Lambda,t-\tau}(w(\bullet + \tau))
    \right] =0
  \end{equation}
  for all $x\in\partial\Lambda$, $t>0$. Therefore, regularity of $\Lambda$ is
  implied by uniform regularity, as it should.
\end{structure}
There are several known conditions implying regularity, see, for example,
\Cite[Section~II.1]{Bas95}, \Cite[Section~2.3]{PoSt78}. Here we only recall
{\bf Poincar\'e's cone condition}, because it may easily be adapted to uniform
regularity. The open set
\begin{equation}\label{2.47}
  C_{r,\beta}(x,u) := 
  \left\{
    y\in\rz^d : 0<|y-x|<r,\, 0< u \cdot (y-x)< |y-x|\cos\beta 
  \right\}
\end{equation}
is called a {\bf finite cone} with vertex at $x\in\rz^d$ in direction
$u\in\rz^d$, $|u|=1$, with opening angle $0<\beta<\frac\pi{2}$ and radius
$r>0$.
\begin{structure}{Proposition}\label{2-12}\structindent\it
  Let $\Lambda\subset\rz^d$ open.  
\structitem\label{2-12-i} If for all $x\in\partial\Lambda$ there is a finite
  cone $C_{r,\beta}(x,u)\subset\rz^d\backslash\Lambda$, $u\in\rz^d$, $|u|=1$,
  $r>0$, $0<\beta<\frac\pi{2}$, then $\Lambda$ is regular.
\structitem\label{2-12-ii} If there are constants $0<\beta<\frac\pi{2}$,
  $r>0$ such that for all $x\in\partial\Lambda$ there is a finite cone
  $C_{r,\beta}(x,u)\subset\rz^d\backslash\Lambda$, $u\in\rz^d$, $|u|=1$, then
  $\Lambda$ is uniformly regular.
\end{structure}
\subsubsection*{Proof:}
The first assertion is proven in \Cite{Bas95} as Proposition~1.13. This may be
seen, using Remark~\ref{2-11-i}.

To show the second part, we use for $x\in\partial\Lambda$ the estimate
\begin{equation}\label{2.48}
  \begin{split}\Ds 
    \E_{x}\!\left[
      \Xi_{\Lambda,t-\tau}(w(\bullet + \tau))
    \right] 
  \le \; & \Ds 
    \E_{x}\!\left[
      \Xi_{\rz^d\backslash\overline{C_{r,\beta}(x,u)},t-\tau}
      (w(\bullet + \tau))
    \right] 
  \\ \Ds  = \; & \Ds 
    \E_{0}\!\left[
      \Xi_{\rz^d\backslash\overline{C_{r,\beta}(0,\overline{u})},t-\tau}
      (w(\bullet + \tau))
    \right] ,
  \end{split}
\end{equation}
where the equality follows from the rotation and translation invariance of
Brownian motion and $\overline{u}$ is any given unit vector. By the first part
of the proposition, $\rz^d\backslash\overline{C_{r,\beta}(0,\overline{u})}$ is
regular. Thus the right-hand side of \xref{2.48}, which is independent of
$x\in\rz^d$, tends to $0$ as $\tau\downarrow 0$ due to Remark~\ref{2-11-ii}.
This proves \xref{2.42}, whence uniform regularity.\halmos
\Section{Continuity of the semigroup in its parameter}
\label{3}
The following result is a straightforward generalization of
\Cite[Proposition~3.2]{Car79} to non-zero magnetic fields and
$\Lambda\subseteq\rz^d$.
\begin{structure}{Theorem}\label{3-1}\structindent\it
  Let $A\in\H_{\text{loc}}(\rz^d)$, $V\in\K_{\pm}(\rz^d)$ and
  $\Lambda\subseteq\rz^d$ open. Moreover, let $1\le p<\infty$ be finite. Then
  the semigroup
  \begin{equation}\label{3.1}
    \left\{
      \e^{-tH_\Lambda(A,V)} :\:
      \L^p(\Lambda) \rightarrow  \L^p(\Lambda)
    \right\}_{t\ge 0}
  \end{equation}
  as defined in Section~\ref{2} is {\bf strongly continuous}, that is,
  \begin{equation}\label{3.2}
    \lim_{t'\to t}
    \left\|
      \left(
        \e^{-tH_\Lambda(A,V)} - \e^{-t^\prime H_\Lambda(A,V)}
      \right)\psi
    \right\|_p
    = 0
  \end{equation}
  for all $\psi\in\L^p(\Lambda)$ and all $t\ge 0$.
\end{structure}
\subsubsection*{Proof:}
Due to the semigroup property \xref{2.37} for $p=q=r$ we may assume $0\le t$,
$t^\prime\le 1$. Then the norm in \xref{3.2} is bounded from above by
\begin{equation}\label{3.3}
   \left(
      \sup_{0\le t\le 1}
      \left\|
            \e^{-tH_\Lambda(A,V)}
      \right\|_{p,p}
   \right)
   \left\|
      \left(
         \e^{-|t-t^\prime|H_\Lambda(A,V)} - 1
      \right)\psi
   \right\|_p.
\end{equation}
Using \xref{2.40} it is therefore sufficient to show \xref{3.2} for $t=0$.
Since \xref{3.2} holds in the free case $A=0$, $V=0$ and $\Lambda=\rz^d$, it is
enough to establish
\begin{equation}\label{3.4}
   \lim_{t\downarrow 0}
   \left\|
      D_{t,t}\psi
   \right\|_p
   =0.
\end{equation}
Here we have made use of the abbreviation
\begin{equation}\label{3.5}
  D_{t,\tau} \defeq
  \e^{-\tau H_{\rz^d}(0,0)}  \e^{-(t-\tau)H_\Lambda(A,V)} -
  \e^{-tH_\Lambda(A,V)}
\end{equation}
where $0\le\tau\le t$. The Feynman-Kac-It\^o formula \xref{2.34} and the Jensen
inequality $|\E_{x}[\bullet]|^p\le\E_{x}[|\!\bullet\!|^p]$ give
\begin{equation}\label{3.6}
   \left|(D_{t,t}\psi)(x)\right|^p \le
   \E_{x}\!\left[\,
      \left|
         1-\e^{-S_t(A,V|w)}\,\Xi_{\Lambda,t}(w)
      \right|^p\,
      |\psi(w(t))|^p\,
   \right].
\end{equation}
Exploiting that $S_t(A,V|w)$ turns into its complex conjugate under time
reversal of Brownian motion, \xref{3.6} leads upon integration over
$x\in\Lambda$ to
\begin{equation}\label{3.7}
   \left(
      \left\|D_{t,t}\psi\right\|_p
   \right)^p \le
   \int_\Lambda
      \E_{x}\!\left[\,
         \left|
            1-\e^{-S_t(A,V|w)}\,\Xi_{\Lambda,t}(w)
         \right|^p\,
      \right]\,
      |\psi(x)|^p\,
   \,\d x.
\end{equation}
Using $|z-z'|^p\le 2^p(|z|^p+|z'|^p)$ for $z,z'\in\cz$, we obtain
\begin{equation}\label{3.8}
   \begin{split}\Ds 
      \left(
         \left\|D_{t,t}\psi\right\|_p
      \right)^p       
   \le \; & \Ds 2^p\,
      \int_\Lambda
         \E_{x}\!\left[\,
            \left|
               1-\e^{-S_t(A,V|w)}
            \right|^p\,
         \right]\,
         |\psi(x)|^p\,
      \,\d x.
   \\ \Ds  & \Ds \hbox{} + 2^p\,
      \int_\Lambda
         \E_{x}\!\left[\,
            1-\Xi_{\Lambda,t}(w)
         \right]\,
         |\psi(x)|^p\,
      \,\d x
   \end{split}
\end{equation}
and employing additionally $-V\le V^-$ and $t\le 1$ we get
\begin{equation}\label{3.9}
   \left|
      1-\e^{-S_t(A,V|w)}
   \right|^p
   \le
   2^p\left(
      1+\e^{-S_1(0,-pV^-|w)}
   \right).
\end{equation}

The action $S_t(A,V|w)$ vanishes for all $x=w(0)$ almost surely as 
$t\downarrow 0$ due to Remark~\ref{2-8-ii}. Moreover,
\begin{equation}\label{3.10}\!\!\!\!\!\!\!\!\!\!\!\!\!\!\!\!
   \int_\Lambda
      \E_{x}\!\left[\,
         1+\e^{-S_1(0,-pV^-|w)}
      \right]\,
      |\psi(x)|^p
   \,\d x
   \le
   \left(
      1+\left\|\e^{-H_{\rz^d}(0,-pV^-)}\right\|_{\infty,\infty}
   \right)\:
   \left( \|\psi\|_p \right)^p
   <\infty\!\!\!\!\!\!\!\!\!\!\!
\end{equation}
by \xref{2.34} and \xref{2.40}. Hence the dominated-convergence theorem implies
that the first integral on the right-hand side of \xref{3.8} vanishes as
$t\downarrow 0$. In order to show that the second integral vanishes too, we
claim
\begin{equation}\label{3.11}
   \lim_{t\downarrow 0} \sup_{x\in \Lambda_r}
   \E_{x}\!\left[\,
       1-\Xi_{\Lambda,t}(w)
   \right] = 0
\end{equation}
for all $r>0$. Here
\begin{equation}\label{3.12}
  \Lambda_r \defeq 
  \left\{
    x\in\Lambda: |x-y|>r \text{ for all } y\in\partial\Lambda
  \right\}
\end{equation}
denotes the set of points well inside $\Lambda$. In fact, one has 
\begin{equation}\label{3.13}
   \sup_{x\in\Lambda_r}
   \E_{x}\!\left[\,
       1-\Xi_{\Lambda,t}(w)
   \,\right]
   \le
   \P_{0}\!\left\{\,
       \sup_{0< s\le t}\, |w(s)| \ge r
   \,\right\}
\end{equation}
and the right-hand side vanishes as $t\downarrow 0$ due to L\'evy's maximal
inequality \Cite[Equation~(7.6')]{Sim79a}.  This completes the proof of
\xref{3.4}.\halmos
\begin{structure}{Remarks}\label{3-2}\structindent
\structitem\label{3-2-i} Even for $A=0$ and $\Lambda=\rz^d$ Theorem~\ref{3-1}
  is slightly different from \Cite[Proposition~3.2]{Car79} as can be inferred
  from \xref{2.11}. For $A=0$ and $\Lambda\subseteq\rz^d$ with finite Lebesgue
  measure Theorem~\ref{3-1} is contained in \Cite[Theorem~3.17]{ChZh95}.
\structitem\label{3-2-ii} As we will see in the next section, the set
  $\e^{-tH_\Lambda(A,V)}\L^{\infty}(\Lambda)$ contains only continuous
  functions if $A\in\H_{\text{loc}}(\rz^d)$, $V\in\K_{\pm}(\rz^d)$ and $t>0$.
  Therefore, the semigroup $\left\{\e^{-tH_\Lambda(A,V)} : \L^{\infty}(\Lambda)
    \rightarrow \L^{\infty}(\Lambda) \right\}_{t\ge 0}$ is not strongly
  continuous for all pairs
  $(A,V)\in\H_{\text{loc}}(\rz^d)\times\K_{\pm}(\rz^d)$, see also
  \Cite[Remark~3.4]{Car79}. However, for $\Lambda=\rz^d$, consider the closed
  subspace $\C_\infty(\rz^d)$ of $\L^\infty(\rz^d)$ consisting of continuous
  functions vanishing at infinity. A slight modification of the proof of
  \xref{4.6} below shows that the semigroup maps $\C_\infty(\rz^d)$ into
  itself.  Moreover, with somewhat more effort it can be shown that this
  restriction yields a strongly continuous semigroup, confer
  \Cite[Theorem~3.17]{ChZh95}.
\end{structure}
\Section{Continuity of the image functions of the semigroup}
\label{4}
In this section we prove that the operator $\e^{-tH_\Lambda(A,V)}$ is smoothing
in the sense that it maps $\L^p(\Lambda)$ into the set $\C(\Lambda)$ of
complex-valued continuous functions on $\Lambda$.

For the reader's convenience we recall that a family ${\cal F}$ of functions
$f:\Lambda\to\cz$ is called {\bf equicontinuous}, if
\begin{equation}\label{4.1}
   \lim_{x'\to x} \, \sup_{f\in{\cal F}}\,  |f(x)-f(x')| = 0
\end{equation}
for all $x\in\Lambda$ and it is called {\bf uniformly equicontinuous}, if
\begin{equation}\label{4.2}
   \lim_{r\downarrow 0}\, \sup_{x,x'\in\Lambda,\, |x-x'|<r} \,
   \sup_{f\in{\cal F}}\, |f(x)-f(x')| = 0,
\end{equation}
see, for example, \Cite[Section~I.6]{ReSi80}. 

\begin{structure}{Theorem}\label{4-1}\structindent\it
   Let $A\in\H_{\text{loc}}(\rz^d)$, $V\in\K_{\pm}(\rz^d)$ and
   $\Lambda\subseteq\rz^d$ open. Then
   \begin{equation}\label{4.3}
      \e^{-tH_\Lambda(A,V)}\,\L^p(\Lambda) 
      \subseteq \L^q(\Lambda)\cap\C(\Lambda),
      \quad t>0,\;1\le p\le q\le \infty,
   \end{equation}
   and for fixed $t>0$, $1\le p \le \infty$ the family
   \begin{equation}\label{4.4}
      \left\{
         \e^{-tH_\Lambda(A,V)}\,\psi : 
         \psi \in \L^p(\Lambda),\; \|\psi\|_p \le 1
      \right\}
   \end{equation}
   of functions on $\Lambda$ is equicontinuous. Furthermore, the right-hand
   side of \xref{2.34} gives the continuous representative of 
   \begin{equation}\label{4.5}
      x \mapsto
      \left(\e^{-tH_\Lambda(A,V)}\psi\right)\!(x)
      , \quad x\in\Lambda.
   \end{equation}
   Finally,
   \begin{equation}\label{4.6}
      \lim_{|x|\to\infty}
      \left(\e^{-tH_\Lambda(A,V)}\,\psi\right)\!(x)
      =0
   \end{equation}
   for all $\psi\in\L^p(\Lambda)$ with finite $1\le p<\infty$ and all
   $t>0$.
\end{structure}
\subsubsection*{Proof:}
Using the boundedness \xref{2.40} and the semigroup property \xref{2.37} with
$q=r=\infty$ it is sufficient to prove that \xref{4.4} is an equicontinuous
family for $p=\infty$ in order to get both \xref{4.3} and the equicontinuity
for all $1\le p\le\infty$. Since this holds in the free case $A=0$, $V=0$ and
$\Lambda=\rz^d$, one has due to \xref{2.40} that
\begin{equation}\label{4.7}
   \left\{
      \e^{-\tau H_{\rz^d}(0,0)}\,\e^{-(t-\tau)H_\Lambda(A,V)}\,\psi : 
         \psi \in \L^\infty(\Lambda),\; \|\psi\|_\infty \le 1
   \right\}
\end{equation}
is an equicontinuous family for all $0<\tau\le t$.  Therefore, it is enough to
show
\begin{equation}\label{4.8}
   \lim_{\tau\downarrow 0} \, \esssup_{x\in K} \,
   \sup_{\|\psi\|_\infty\le 1}
   \left|
      (D_{t,\tau}\psi)(x)
   \right|
   =0
\end{equation}
for all compact $K\subset\Lambda$.  To this end, we represent the image of
$\psi$ by the operator difference \xref{3.5} as
\begin{equation}\label{4.9}
   \begin{split} \Ds
      (D_{t,\tau}\psi)\,(x) = \E_{x}\!\bigl[\,
   & \Ds
      \e^{S_\tau(A,V|w)-S_t(A,V|w)}
   \\ \Ds & \Ds \hbox{} \times  
      \left(
         \Xi_{\Lambda,t-\tau}(w(\tau+\bullet))
         -\e^{-S_\tau(A,V|w)}\,\Xi_{\Lambda,t}(w)
      \right)\,
      \psi(w(t))\,
      \bigr]
   \end{split}
\end{equation}
where we have used the Feynman-Kac-It\^o formula \xref{2.34}, the additivity of
the integrals in the action \xref{2.33} and the Markov property of the Brownian
motion $w$.  We use the right-hand side of \xref{4.9} to give meaning to
$(D_{t,\tau}\psi)\,(x)$ for all $x\in\rz^d$. Then, in order to show that the
right-hand side of \xref{2.34} defines the continuous representative of
\xref{4.5}, it is sufficient to establish
\begin{equation}\label{4.10}
   \lim_{\tau\downarrow 0} \, \sup_{x\in K} \,
   \sup_{\|\psi\|_\infty\le 1}
   \left|
      (D_{t,\tau}\psi)\,(x)
   \right|
   =0
\end{equation}
for all compact $K\subset\Lambda$.

The triangle inequality in combination with
\begin{equation}\label{4.11}
   S_\tau(0,V|w) - S_t(0,V|w)
   \le -S_t(0,-V^-|w)
\end{equation}
yields
\begin{equation}\label{4.12}\!\!\!\!\!\!\!\!
   \begin{split} \Ds
      \left|(D_{t,\tau}\psi)\,(x)\right| \le
      \E_{x}\!\bigl[\,
   & \Ds
         \e^{-S_t(0,-V^-|w)}
   \\ \Ds & \Ds \hbox{} \times  
         \left|
            \Xi_{\Lambda,t-\tau}(w(\tau+\bullet))
            -\e^{-S_\tau(A,V|w)}\,\Xi_{\Lambda,t}(w)
         \right|\,
      \left|\psi(w(t))\right|\,
      \bigr].
   \end{split}
\end{equation}
By $\left|\psi(w(t))\right|\le\|\psi\|_\infty$ and the Cauchy-Schwarz
inequality one arrives at
\begin{equation}\label{4.13}\!\!\!\!\!\!\!\!
   \begin{split} \Ds
      \left|(D_{t,\tau}\psi)\,(x)\right|^2 \le\;
   & \Ds
      \left( \|\psi\|_\infty \right)^2 \,
      \E_{x}\!\left[\e^{-S_t(0,-2V^-|w)}\right]
   \\ \Ds & \Ds \hbox{} \times  
      \biggl(\,
         \E_{x}\!\left[\,
            \Xi_{\Lambda,t-\tau}(w(\tau+\bullet))
            -\Xi_{\Lambda,t}(w)
         \,\right]
   \\ \Ds & \Ds \phantom{\hbox{}\times\biggl(\,}  
         +
         \E_{x}\!\left[\,
            \left|
              1-\e^{-S_\tau(A,V|w)}
            \right|^2\,
         \right]
      \,\biggr).
   \end{split}
\end{equation}
Since
\begin{equation}\label{4.14}\!\!\!\!\!\!\!\!\!\!\!\!\!\!\!\!
   \begin{split} \Ds
      \E_{x}\!\left[\,
         \Xi_{\Lambda,t-\tau}(w(\tau+\bullet))
         -\Xi_{\Lambda,t}(w)
      \,\right]
   =\; & \Ds
      \E_{x}\!\left[\,
         \Xi_{\Lambda,t-\tau}(w(\tau+\bullet))
         \left(1-\Xi_{\Lambda,\tau}(w)\right)
      \,\right]
   \\ \Ds \le\; & \Ds
      \E_{x}\!\left[\,
         1-\Xi_{\Lambda,\tau}(w)
      \,\right],
   \end{split}
\end{equation}
\xref{4.10} follows from \xref{3.11} and Lemmas~\ref{C.2}, \ref{C-5}. This
completes the proof of \xref{4.3} and the equicontinuity and identifies the
continuous representative.

Finally, for the proof of \xref{4.6} we may assume without loss of generality
$1<p<\infty$ due to \xref{2.40} and the semigroup property. Then \xref{4.6}
follows from H\"older's inequality
\begin{equation}\label{4.15}\!\!\!\!\!\!\!\!\!\!\!\!\!\!\!\!
   \left|\left(\e^{-t H_\Lambda(A,V)}\psi\right)\!(x)\right|
   \le
   \left(\,\left\|   
      \e^{-tH_{\rz^d}(0,p'V)} 
   \,\right\|_{\infty,\infty}\,\right)^{1/p'} \,
   \biggl(\,
      \left(\e^{-t H_{\rz^d}(0,0)}\left|\psi\right|^p\right)\!(x)
   \,\biggr)^{1/p}\!\!\!\!\!\!\!\! 
\end{equation}
where $p'\defeq p/(p-1)<\infty$.\halmos
\begin{structure}{Remarks}\label{4-2}\structindent
\structitem\label{4-2-i} The assertion \xref{4.3} reduces for $A=0$,
  $V\in\K_{\pm}(\rz^d)$ and $\Lambda=\rz^d$ to \Cite[Corollary~B.3.2]{Sim82}.
  A related result, also for $A=0$ and $\Lambda=\rz^d$, is
  \Cite[Propositions~3.1,~3.3]{Car79}.  Our proof is patterned after that of
  \Cite[Lemma~3.2]{Car79} or \Cite[Propositions~3.11,~3.12]{ChZh95} and, in
  contrast to the strategy in \Cite{Sim82}, does not use Propositions~\ref{2-3}
  and \ref{2-6}.
\structitem\label{4-2-ii} For $V=0$, $\nabla\cdot A=0$ and 
  $A^2\in\K_{\text{loc}}(\rz^d)$ \Cite[Theorem~3.1]{Smi89} asserts that
   \begin{equation}\label{4.16}
      \e^{-tH_{\rz^d}(A,0)}
      \left( \L^\infty(\rz^d)\cap\L^2(\rz^d) \right)
      \subseteq \L^\infty(\rz^d)\cap\C(\rz^d)
      ,\quad t>0.
   \end{equation}
   The proof given there, however, applies for all $A^2\in\K(\rz^d)$
   but not for all $A^2\in\K_{\text{loc}}(\rz^d)$, because it can happen that
   \begin{equation}\label{4.17}
      \sup_{x\in K}
      \E_{x}\!\left[\int_0^t(A(w(s)))^2\,\d s\right]
      = \infty
   \end{equation}        
   although $A^2\in\K_{\text{loc}}(\rz^d)$ and $K\subset\rz^d$
   compact. Consider the example
   \begin{equation}\label{4.18}
      A(x) = \left(x_2,-x_1\right)\:\e^{|x|^4}
   \end{equation}
   of a vector potential on $\rz^2$.
\end{structure}
Since $H_\Lambda(A,V)$ is equipped with Dirichlet boundary conditions on
$\partial\Lambda$ one would expect that
\begin{equation}\label{4.19}
  \lim_{x\to y}
  \left(\e^{-tH_\Lambda(A,V)}\psi\right)\!(x) =0
  , \quad y\in\partial\Lambda,
\end{equation}
for a sufficiently nice boundary of $\Lambda$. 
\begin{structure}{Theorem}\label{4-3}\structindent\it
  Let $A\in\H_{\text{loc}}(\rz^d)$, $V\in\K_{\pm}(\rz^d)$ and
  $\Lambda\subset\rz^d$ regular. Then \xref{4.19} holds for all
  $\psi\in\L^p(\Lambda)$, $1\le p\le\infty$ and $t>0$. Furthermore, \xref{4.4}
  is an equicontinuous family of functions on $\overline{\Lambda}$ for all
  $1\le p\le\infty$, $t>0$.
\end{structure}
\subsubsection*{Proof:}
The semigroup property \xref{2.37} and \xref{2.40} ensure that it is sufficient
to check the case $p=\infty$. The triangle inequality, the Cauchy-Schwarz
inequality and \xref{2.34} imply
\begin{equation}\label{4.20}
  \left|\left(\e^{-t H_\Lambda(A,V)}\psi\right)\!(x)\right|
  \le
  \left\| \psi \right\|_\infty \,
  \left(\,\left\|   
    \e^{-tH_{\rz^d}(0,2V)} 
  \,\right\|_{\infty,\infty}\,\right)^{1/2} \,
  \bigl(\,
    \E_{x}\!\left[\,\Xi_{\Lambda,t}(w)\,\right]
  \,\bigr)^{1/2}
\end{equation}
for almost all $x\in\rz^d$. Now the assertions follow from Theorem~\ref{4-1},
\xref{2.40} and Lemma~\ref{C-7}.\halmos
Not surprisingly, uniform continuity of the image functions can be achieved for
sufficiently regular $\Lambda\subseteq\rz^d$ by imposing global regularity
conditions for the potentials, thereby, however, possibly excluding physically
relevant cases, confer Remark~\ref{2-5-i}.
\begin{structure}{Theorem}\label{4-4}\structindent\it
  Let $A\in\H(\rz^d)$, $V\in\K(\rz^d)$ and $\Lambda\subseteq\rz^d$ uniformly
  regular. Then \xref{4.4} is a uniformly equicontinuous family of functions on
  $\overline{\Lambda}$ for all $1\le p\le\infty$ and $t>0$.
\end{structure}
\subsubsection*{Proof:}
Analogously to the reasoning at the beginning of the proof of Theorem~\ref{4-1}
it suffices to check
\begin{equation}\label{4.21}
   \lim_{\tau\downarrow 0} \, \esssup_{x\in\Lambda_r} \,
   \sup_{\|\psi\|_\infty\le 1}\, 
   \left|
      (D_{t,\tau}\psi)\,(x)
   \right|
   =0
\end{equation}
in order to get uniform equicontinuity on $\Lambda_r$, $r>0$. This follows with
the help of \xref{4.13} and \xref{4.14} from \xref{3.11}, Lemma~\ref{C-2} and
Lemma~\ref{C-3}. Using Lemma~\ref{C-7} and the estimate \xref{4.20} we conclude
\begin{equation}\label{4.22}
  \lim_{r\downarrow 0}\, \esssup_{x\in\Lambda\backslash\Lambda_r}\,
   \sup_{\|\psi\|_\infty\le 1}\, 
  \left|\left(\e^{-t H_\Lambda(A,V)}\psi\right)\!(x)\right|
  =0,
\end{equation}
which is sufficient to extend the domain of uniform equicontinuity from
$\Lambda_r$ for $r>0$ to $\overline{\Lambda}$.\halmos
\begin{structure}{Remark}\label{4-5}\structindent
  Proposition~3.1 in \Cite{Smi92} is a special case of Theorem~\ref{4-4} and
  \xref{4.6}.
\end{structure}
\Section{Continuity of the semigroup in the potentials}
\label{5}
As a motivation for this section consider the simple example
\begin{equation}\label{5.1}
  A_h(x) \defeq \left( 0, x_1 h \right),\quad h>0,
\end{equation}
of a vector potential on $\rz^2$. Clearly, $A_h$ belongs to
$\H_{\text{loc}}(\rz^2)$ and gives rise to a constant magnetic field of
strength $h$. The related semigroup
\begin{equation}\label{5.2}
   \left\{ \e^{-tH_{\rz^2}(A_h,0)} :\:
   \L^2(\rz^2) \rightarrow \L^2(\rz^2) \right\}_{t\ge 0},
\end{equation}
considered as a function of $h$, is not norm-continuous because
\begin{equation}\label{5.3}
   \limsup_{h\downarrow 0}
   \left\|\,
      \e^{-tH_{\rz^2}(A_h,0)} -
      \e^{-tH_{\rz^2}(0,0)}
   \,\right\|_{2,2}
   > 0
\end{equation}
for all $t>0$. This can be deduced, for example, from
\Cite[Theorem~6.3]{AvHe78}. It reflects the fact that the character of the
energy spectrum of an electrically charged point-mass in the Euclidean plane
changes from purely continuous to pure point, when an arbitrarily low constant
magnetic field, perpendicular to the plane, is turned on.

The following theorem, however, shows that under suitable technical assumptions
the weaker notion of {\bf local-norm-continuity} holds.
\begin{structure}{Theorem}\label{5-1}\structindent\it
   Let $A\in\H_{\text{loc}}(\rz^d)$,
   $\left\{A_m\right\}_{m\in\nz}\subset\H_{\text{loc}}(\rz^d)$, 
   $V\in\K_{\pm}(\rz^d)$ and
   $\left\{V_n\right\}_{n\in\nz}\subset\K_{\pm}(\rz^d)$ such that
   for all compact $K\subset\rz^d$
   \begin{gather}\label{5.4}
         \lim_{m\to\infty}
         \left\|\,
            \left(A-A_m\right)^2
            \,\chi_{_{\scriptstyle K}}
         \,\right\|_{\K(\rz^d)}
         = 0 ,
   \\ \label{5.5}
         \lim_{m\to\infty}
         \left\|\,
            \left(\nabla\cdot A-\nabla\cdot A_m\right)
            \,\chi_{_{\scriptstyle K}}
         \,\right\|_{\K(\rz^d)}
         = 0
   \end{gather}
   and
   \begin{gather}\label{5.6}
         \lim_{n\to\infty}
         \left\|\,
            \left(V-V_n\right)
            \,\chi_{_{\scriptstyle K}}
         \,\right\|_{\K(\rz^d)}
         = 0  ,
   \\ \label{5.7}
      \lim_{\rho\downarrow 0}\sup_{n\in\nz}\sup_{x\in\rz^d}\:
      \int g_\rho(x-y)\:V^-_n(y)\,\d y
      = 0.
   \end{gather}
   Moreover, let $\Lambda\subseteq\rz^d$ open. Then
   \begin{equation}\label{5.8}
      \lim_{m,n\to\infty} \sup_{\tau_1 \le t \le \tau_2}
      \left\|\,
         \chi_{_{\scriptstyle K}}\,
         \left(
            \e^{-tH_\Lambda(A,V)} -
            \e^{-tH_\Lambda(A_m,V_n)}
         \right)
      \,\right\|_{p,q}
      = 0
   \end{equation}
   and
   \begin{equation}\label{5.9}
      \lim_{m,n\to\infty} \sup_{\tau_1 \le t \le \tau_2}
      \left\|\,
         \left(
            \e^{-tH_\Lambda(A,V)} -
            \e^{-tH_\Lambda(A_m,V_n)}
         \right)
         \,\chi_{_{\scriptstyle K}}
      \,\right\|_{p,q}
      = 0
   \end{equation}
   for all compact $K\subset\rz^d$, $0 < \tau_1 \le \tau_2 < \infty$ and
   $1 \le p \le q \le \infty$. Furthermore, for $p=q$ one may allow 
   $\tau_1=0$.
\end{structure}
\subsubsection*{Proof:}
According to the Riesz-Thorin interpolation theorem
\Cite[Theorem~IX.17]{ReSi75} it is enough to prove the theorem for the three
cases $p=q=1$, $p=q=\infty$ and $p=1$, $q=\infty$. Moreover, due to the
self-adjointness of the semigroup, the assertions \xref{5.8} and \xref{5.9} are
equivalent under the combined substitutions $p\mapsto\left(1-\frac
  1p\right)^{-1}$, $q\mapsto\left(1-\frac 1q\right)^{-1}$. In consequence, it
remains to show the following three partial assertions:
\begin{itemize}
\item[{\bf 1)}] \xref{5.8} for $p=q=\infty$, $\tau_1=0$
\item[{\bf 2)}] \xref{5.9} for $p=q=\infty$, $\tau_1=0$
\item[{\bf 3)}] \xref{5.8} for $p=1$, $q=\infty$, $\tau_1>0$.
\end{itemize}
We note that \xref{5.8} and \xref{5.9} in the case $p=q$, $\tau_1=0$ follow
already from the assertions 1) and 2).
\subsubsection*{As to assertion 1)}
Since the Feynman-Kac-It\^o formula \xref{2.34} and the triangle inequality
give
\begin{equation}\label{5.10}
   \begin{split}\Ds & \Ds
      \left\|\,
         \chi_{_{\scriptstyle K}}\,
         \left(
            \e^{-tH_\Lambda(A,V)} -
            \e^{-tH_\Lambda(A_m,V_n)}
         \right)
      \,\right\|_{\infty,\infty}
   \\ \Ds \le\; & \Ds
      \sup_{x\in K}
      \E_{x}\!\left[\,
         \left|
            \e^{-S_t(A,V|w)}
            -\e^{-S_t(A_m,V_n|w)}
        \right|\,
      \right],
   \end{split}
\end{equation}
the assertion follows from Lemma~\ref{C-6}.
\subsubsection*{As to assertion 2)}
Let $B_R$ be the open ball of radius $R>0$ centered about the origin in
$\rz^d$, see \xref{2.1}. Then the diamagnetic inequality \xref{2.38} in
combination with \xref{2.39} and the triangle inequality yields
\begin{equation}\label{5.11}\!\!\!\!\!\!\!\!\!\!\!\!
   \begin{split}\Ds & \Ds
      \left\|\,
         \left(
            \e^{-tH_\Lambda(A,V)} -
            \e^{-tH_\Lambda(A_m,V_n)}
         \right)
         \,\chi_{_{\scriptstyle K}}
      \,\right\|_{\infty,\infty}
    \\ \Ds \le\; &\Ds
      \left\|\,
         \chi_{_{\scriptstyle B_R}}\,
         \left(
            \e^{-tH_\Lambda(A,V)} -
            \e^{-tH_\Lambda(A_m,V_n)}
         \right)
      \,\right\|_{\infty,\infty}
   \\ \Ds  & \Ds
      \hbox{} +
      \left\|\,
         \left( 1-\chi_{_{\scriptstyle B_R}} \right)\,
         \e^{-tH_{\rz^d}(0,V)}
         \,\chi_{_{\scriptstyle K}}
      \,\right\|_{\infty,\infty}
      +
      \left\|\,
         \left( 1-\chi_{_{\scriptstyle B_R}} \right)\,
         \e^{-tH_{\rz^d}(0,V_n)}
         \,\chi_{_{\scriptstyle K}}
      \,\right\|_{\infty,\infty}.
   \end{split}\!\!\!\!\!\!\!\!\!\!\!\!
\end{equation}
Hence assertion 2 follows from assertion 1 provided that
\begin{equation}\label{5.12}
   \lim_{R\to\infty} \sup_{0 \le t \le \tau_2} \sup_{n\in\nz}
   \left\|\,
      \left( 1-\chi_{_{\scriptstyle B_R}} \right)\,
      \e^{-tH_{\rz^d}(0,V_n)} 
      \,\chi_{_{\scriptstyle K}}
   \,\right\|_{\infty,\infty}
   = 0.
\end{equation}
To prove \xref{5.12} we use the Cauchy-Schwarz inequality in the
Feynman-Kac-It\^o formula to obtain
\begin{equation}\label{5.13}\!\!\!\!\!\!\!\!
   \begin{split}\Ds & \Ds
      \left\|\,
         \left( 1-\chi_{_{\scriptstyle B_R}} \right)\,
         \e^{-tH_{\rz^d}(0,V_n)}
         \,\chi_{_{\scriptstyle K}}
      \,\right\|_{\infty,\infty}
      =
      \sup_{x\not\in B_R}
      \E_{x}\!\left[\,
         \e^{-S_t(0,V_n|w)} \:
         \chi_{_{\scriptstyle K}}(w(t))\,
      \right]
   \\ \Ds \le \; & \Ds
      \left(\:
         \left\|\,
            \left( 1-\chi_{_{\scriptstyle B_R}} \right)\,
            \e^{-tH_{\rz^d}(0,0)}
            \,\chi_{_{\scriptstyle K}}
         \,\right\|_{\infty,\infty}
      \right)^{1/2}\;
      \left(\:
         \left\|\,
            \e^{-tH_{\rz^d}(0,2V_n)}
         \,\right\|_{\infty,\infty}
      \right)^{1/2}.
   \end{split}\!\!\!\!\!\!\!\!
\end{equation}
By Lemma~\ref{C-1} the second factor on the right-hand side is uniformly
bounded with respect to $n\in\nz$ and $0 \le t \le \tau_2$. The first factor is
seen to vanish uniformly in $0\le t \le \tau_2$ as $R\to\infty$ by an
elementary calculation.
\subsubsection*{As to assertion 3)}
In a first step we get with the help of the triangle inequality
\begin{equation}\label{5.14}
   \left\|\,
      \chi_{_{\scriptstyle K}}\,
      \left(
         \e^{-2tH_\Lambda(A,V)} -
         \e^{-2tH_\Lambda(A_m,V_n)}
      \right)
   \,\right\|_{1,\infty}
   \le N_1 + N_2 + N_3,
\end{equation}
where
\begin{align}\Ds \label{5.15}\!\!\!\!\!\!\!\!
   N_1 &\Ds \defeq
   \left\|\,
      \chi_{_{\scriptstyle K}}\,
      \left(
         \e^{-tH_\Lambda(A,V)} -
         \e^{-tH_\Lambda(A_m,V_n)}
      \right)\,
      \e^{-tH_\Lambda(A,V)}
   \,\right\|_{1,\infty},
\\ \Ds \label{5.16}\!\!\!\!\!\!\!\!
   N_2 &\Ds \defeq
   \left\|\,
      \chi_{_{\scriptstyle K}}\,
      \e^{-tH_\Lambda(A_m,V_n)}\,
      \chi_{_{\scriptstyle B_R}}\,
      \left(
         \e^{-tH_\Lambda(A,V)} -
         \e^{-tH_\Lambda(A_m,V_n)}
      \right)
   \,\right\|_{1,\infty},
\\ \Ds \label{5.17}\!\!\!\!\!\!\!\!
   N_3 &\Ds \defeq
   \left\|\,
      \chi_{_{\scriptstyle K}}\,
         \e^{-tH_\Lambda(A_m,V_n)}\,
         \left(
            1 - \chi_{_{\scriptstyle B_R}}\,
         \right)
      \left(
         \e^{-tH_\Lambda(A,V)} -
         \e^{-tH_\Lambda(A_m,V_n)}
      \right)
   \,\right\|_{1,\infty}.
\end{align}
In a second step we will repeatedly use the inequality
\begin{equation}\label{5.18}
   \left\|\, XY \,\right\|_{p,r}
   \le
   \left\|\, X \,\right\|_{q,r} \: \left\|\, Y \,\right\|_{p,q}
\end{equation}
for bounded operators $Y:\L^p(\rz^d)\rightarrow\L^q(\rz^d)$ and
$X:\L^q(\rz^d)\rightarrow\L^r(\rz^d)$. By \xref{5.18} and \xref{2.40} we get
\begin{equation}\label{5.19}
   N_1 \le
   \left\|\,
      \chi_{_{\scriptstyle K}}\,
      \left(
         \e^{-tH_\Lambda(A,V)} -
         \e^{-tH_\Lambda(A_m,V_n)}
      \right)\,
   \,\right\|_{\infty,\infty}\;
   \left\|\,
      \e^{-tH_{\rz^d}(0,V)}
   \,\right\|_{1,\infty}.
\end{equation}
By employing additionally the self-adjointness of the semigroup we obtain
\begin{equation}\label{5.20}
   N_2 \le
   \left\|\,
      \e^{-tH_{\rz^d}(0,V_n)}\,
   \,\right\|_{1,\infty}\;
   \left\|\,
      \left(
         \e^{-tH_\Lambda(A,V)} -
         \e^{-tH_\Lambda(A_m,V_n)}
      \right)\,
      \chi_{_{\scriptstyle B_R}}\,
   \,\right\|_{\infty,\infty}
\end{equation}
and similarly
\begin{equation}\label{5.21}
   \begin{split}\Ds
      N_3 \le \;
   & \Ds
      \left\|\,
         \chi_{_{\scriptstyle K}}\,
         \e^{-tH_{\rz^d}(0,V_n)}\,
         \left(
            1 - \chi_{_{\scriptstyle B_R}}\,
        \right)
      \,\right\|_{\infty,\infty}\;
    \\ \Ds  &\Ds \hbox{} \times
      \left(\,
         \left\|\,
            \e^{-tH_{\rz^d}(0,V)}
         \,\right\|_{1,\infty}
         +
         \left\|\,
            \e^{-tH_{\rz^d}(0,V_n)}
         \,\right\|_{1,\infty}
      \,\right).     
   \end{split}
\end{equation}
Hence assertion~3 follows from assertions~1 and~2 together with Lemma~\ref{C-1}
and an asymptotic relation analogous to \xref{5.12}.\halmos
\begin{structure}{Remarks}\label{5-2}\structindent
\structitem\label{5-2-i} According to \xref{2.29} the analytic condition
  \xref{5.7} is equivalent to the probabilistic condition
  \begin{equation}\label{5.22}
    \lim_{t\downarrow 0}\,\sup_{n\in\nz}\,\sup_{x\in\rz^d}\,
    \E_{x}\!\left[\int_0^tV_n^-(w(s))\,\d s\right]
    = 0.
  \end{equation}
\structitem\label{5-2-ii} Theorem~\ref{5-1} is a generalization of
  \Cite[Theorem~B.10.2]{Sim82} both to non-zero magnetic fields and
  $\Lambda\subseteq\rz^d$.  Even for $A=A_m=0$ and $\Lambda=\rz^d$ the result
  is slightly stronger than that of \Cite[Theorem~B.10.2]{Sim82}.
  Nevertheless, we followed a similar strategy for the proof.
\structitem\label{5-2-iii} Propositions~\ref{2-3} and \ref{2-6} imply that
  for given $A\in\H_{\text{loc}}(\rz^d)$ and $V\in\K_{\pm}(\rz^d)$ one can find
  sequences $\{A_m\}_{m\in\nz}\subset\left(\C^\infty_0(\rz^d)\right)^d$ and
  $\left\{V_n\right\}_{n\in\nz}\subset\C^\infty_0(\rz^d)$ obeying the
  hypotheses of Theorem~\ref{5-1}.
\end{structure}

Since the notion of local-norm-continuity occurring in Theorem~\ref{5-1} seems
to us less common, it may be worth noting that it implies strong continuity of
the semigroup in the potentials under the additional condition $p<\infty$.
\begin{structure}{Corollary}\label{5-3}\structindent\it
  Let $A\in\H_{\text{loc}}(\rz^d)$,
  $\left\{A_m\right\}_{m\in\nz}\subset\H_{\text{loc}}(\rz^d)$,
  $V\in\K_{\pm}(\rz^d)$ and
  $\left\{V_n\right\}_{n\in\nz}\subset\K_{\pm}(\rz^d)$ obey \xref{5.4}
  -- \xref{5.7} for all compact $K\subset\rz^d$. Moreover, let
  $\Lambda\subseteq\rz^d$ open. Then
   \begin{equation}\label{5.23}
      \lim_{m,n\to\infty} \sup_{\tau_1 \le t \le \tau_2}
      \left\|\,
         \left(
            \e^{-tH_\Lambda(A,V)} -
            \e^{-tH_\Lambda(A_m,V_n)}
         \right)
         \psi
      \,\right\|_{q}
      = 0
   \end{equation}
   for all $0<\tau_1\le\tau_2<\infty$, $\psi\in\L^p(\Lambda)$, $1\le
   p<\infty$ being finite, and $1\le p\le q\le\infty$. Furthermore, for
   $p=q<\infty$ one may allow $\tau_1=0$.
\end{structure}
\subsubsection*{Proof:}
We repeatedly use the triangle inequality to achieve the estimate
\begin{equation}\label{5.24}
   \begin{split} \Ds & \Ds
      \left\|\,
         \left(
            \e^{-tH_\Lambda(A,V)} -
            \e^{-tH_\Lambda(A_m,V_n)}
         \right)
         \psi
      \,\right\|_{q}
   \\ \Ds \le\; & \Ds
      \left\|\,
         \left(
            \e^{-tH_\Lambda(A,V)} -
            \e^{-tH_\Lambda(A_m,V_n)}
         \right)
         \chi_{_{\scriptstyle B_R}}
         \,\psi
      \,\right\|_{q}
   \\ \Ds  &\Ds \hbox{} +
      \left(
         \left\|\,
            \e^{-tH_{\rz^d}(0,V)}
         \right\|_{p,q}
         +
         \left\|\,
            \e^{-tH_{\rz^d}(0,V_n)}
         \right\|_{p,q}
      \right)
      \left\|\,
         \left( 1 - \chi_{_{\scriptstyle B_R}} \right) \psi
      \right\|_{p}
   \end{split}
\end{equation}
valid for any $R>0$. The right-hand side of the estimate is uniformly bounded
in $n,m\in\nz$, $\tau_1\le t\le\tau_2$, due to \xref{2.40} and Lemma~\ref{C-1},
respectively. The proof is therefore accomplished with the help of
Theorem~\ref{5-1} and the fact that
\begin{equation}\label{5.25}
   \lim_{R\to\infty}
   \left\|\,
      \left( 1 - \chi_{_{\scriptstyle B_R}} \right) \psi
   \right\|_{p}
   =0,
\end{equation}
whenever $\psi\in\L^p(\Lambda)$, $1\le p<\infty$.\halmos
\begin{structure}{Remark}\label{5-4}\structindent
  For $p=2$ Corollary~\ref{5-3} is a special case of \Cite[Theorem~2.8]{LiMa97}
  as can be seen from Remark~\ref{2-2-ii}.
\end{structure}

If one defers norm-continuity in the potentials instead of
local-norm-continuity, one has to make stronger assumptions as indicated in the
beginning of this section.
\begin{structure}{Theorem}\label{5-5}\structindent\it
   Let $\left\{A_m\right\}_{m\in\nz}\subset\H(\rz^d)$
   and $\left\{V_n\right\}_{n\in\nz}\subset\K(\rz^d)$ such that
   \begin{gather}\label{5.26}
      \lim_{m\to\infty}
      \left\|\,
         \left(A_m\right)^2
      \,\right\|_{\K(\rz^d)}
      = 0,
   \\ \label{5.27}
      \lim_{m\to\infty}
      \left\|\,
         \nabla\cdot A_m
      \,\right\|_{\K(\rz^d)}
      = 0
   \end{gather}
   and
   \begin{equation}\label{5.28}
      \lim_{n\to\infty}
      \left\|\,
         V_n
      \,\right\|_{\K(\rz^d)}
      = 0.
   \end{equation}
   Moreover, let $A\in\H_{\text{loc}}(\rz^d)$, $V\in\K_{\pm}(\rz^d)$ and
   $\Lambda\subseteq\rz^d$ open. Then
   \begin{equation}\label{5.29}
      \lim_{m,n\to\infty} \sup_{\tau_1 \le t \le \tau_2}
      \left\|\,
         \left(
            \e^{-tH_\Lambda(A,V)} -
            \e^{-tH_\Lambda(A+A_m,V+V_n)}
         \right)
      \,\right\|_{p,q}
      = 0
   \end{equation}
   for all $0 < \tau_1 \le \tau_2 < \infty$ and $1 \le p \le q \le \infty$.
   Furthermore, for $p=q$ one may allow $\tau_1=0$.
\end{structure}
\subsubsection*{Proof:}
According to the Riesz-Thorin interpolation theorem
\Cite[Theorem~IX.17]{ReSi75} and the self-adjointness of the semigroup it is
enough to prove \xref{5.29} for the two cases $p=q=\infty$, $\tau_1=0$ and
$p=1$, $q=\infty$, $\tau_1>0$.

Since the Feynman-Kac-It\^o formula \xref{2.34}, the triangle inequality and
the Cauchy-Schwarz inequality give
\begin{equation}\label{5.30}
   \begin{split}\Ds & \Ds
      \left\|\,
         \e^{-tH_\Lambda(A,V)} -
         \e^{-tH_\Lambda(A+A_m,V+V_n)}
      \,\right\|_{\infty,\infty}
   \\ \Ds \le\; & \Ds
      \left(\left\|\,
         \e^{-tH_{\rz^d}(0,2V)}
      \,\right\|_{\infty,\infty}\right)^{1/2}
      \left(
         \sup_{x\in\rz^d}
         \E_{x}\!\left[\,
            \left|
               1-\e^{-S_t(A_m,V_n|w)}
           \right|^2\,
         \right]
      \right)^{1/2},
   \end{split}
\end{equation}
the case $p=q=\infty$, $\tau_1=0$ follows from \xref{2.40} and Lemma~\ref{C-4}.

By reasoning in a similar way to the proof of assertion 3 in the proof of
Theorem~\ref{5-1} we get the estimate
\begin{equation}\label{5.31}
   \begin{split}\Ds & \Ds
      \left\|\,
         \e^{-2tH_\Lambda(A,V)} -
         \e^{-2tH_\Lambda(A+A_m,V+V_n)}
      \,\right\|_{1,\infty}
   \\ \Ds \le\; & \Ds
      \left\|\,
         \e^{-tH_\Lambda(A,V)} -
         \e^{-tH_\Lambda(A+A_m,V+V_n)}
      \,\right\|_{\infty,\infty}
   \\ \Ds  & \Ds \hbox{} \times
      \left(
         \left\|\,
            \e^{-tH_{\rz^d}(0,V)}
         \,\right\|_{1,\infty}
         +
         \left\|\,
            \e^{-tH_{\rz^d}(0,V+V_n)}
         \,\right\|_{1,\infty}
      \right).
   \end{split}
\end{equation}
The second factor on the right-hand side of \xref{5.31} is uniformly bounded
with respect to $n\in\nz$ and $\tau_1\le t\le\tau_2$ due to \xref{2.40} and
Lemma~\ref{C-1}, the latter being applicable because
$\left\{V_n\right\}_{n\in\nz}\subset\K(\rz^d)$ and \xref{5.28} together with
the definition~\xref{2.10} imply \xref{5.7}. Thus the case $p=1$, $q=\infty$,
$\tau_1>0$ follows with the help of the preceding case.\halmos
\begin{structure}{Remarks}\label{5-6}\structindent
\structitem\label{5-6-i}
  Theorem~\ref{5-5} is a generalization of \Cite[Theorem~B.10.1]{Sim82}.
\structitem\label{5-6-ii} In analogy to Remark~\ref{5-2-iii} one can find for
  given $A\in\H_{\text{loc}}(\rz^d)$ and $V\in\K_{\pm}(\rz^d)$ sequences
  $\{A+A_m\}_{m\in\nz}\subset\left(\C^\infty(\rz^d)\right)^d$ and
  $\left\{V+V_n\right\}_{n\in\nz}\subset\C^\infty(\rz^d)$ obeying the
  hypotheses of Theorem~\ref{5-5}. In contrast to Remark~\ref{5-2-iii} it is in
  general wrong to replace here the set $\C^\infty(\rz^d)$ of arbitrarily often
  differentiable functions by $\C_0^\infty(\rz^d)$, since the Kato-norm closure
  of $\C_0^\infty(\rz^d)$ is a proper subspace of $\K(\rz^d)$, see
  \Cite[Proposition~5.5]{Voi86}.
\end{structure}
\Section{Continuity of the integral kernel of the semigroup}
\label{6}
From the Dunford-Pettis theorem \Cite[Theorem~46.1]{Tre67},
\Cite[Corollary~2.14]{CyFr87} it follows with the help of \xref{2.40} that the
operator $\e^{-tH_\Lambda(A,V)}:\L^p(\Lambda)\to\L^q(\Lambda)$, $t>0$, $1\le
p\le q\le\infty$, $A\in\H_{\text{loc}}(\rz^d)$, $V\in\K_{\pm}(\rz^d)$,
$\Lambda\subseteq\rz^d$ open, has an {\bf integral kernel}
$k_t:\Lambda\times\Lambda\to\cz$ in the sense that
\begin{equation}\label{6.1}
   \left(\e^{-tH_\Lambda(A,V)}\,\psi\right)(x)
   =
   \int_\Lambda k_t(x,y) \, \psi(y) \,\d y
\end{equation}
for all $\psi\in\L^p(\Lambda)$ and almost all $x\in\Lambda$.  Furthermore, the
integral kernel is bounded according to
\begin{equation}\label{6.2}
  \esssup_{x,y\in\Lambda}\, \left| k_t(x,y) \right|
  =  
  \left\|\, \e^{-tH_\Lambda(A,V)} \,\right\|_{1,\infty} .
\end{equation}
The existence of an integral kernel can also be inferred from the
Feynman-Kac-It\^o formula \xref{2.34} by conditioning the Brownian motion to
arrive at $y$ at time $t$. The resulting representation
\begin{equation}\label{6.3}
   k_t(x,y)
   =
   \left(2\pi t\right)^{-d/2} \, \e^{-(x-y)^2/2t} \:
   \E_{x}\!\left[\left.
       \e^{-S_t(A,V|w)}\,\Xi_{\Lambda,t}(w)
   \right|
       w(t)=y
   \right]
\end{equation}
holds for almost all pairs $(x,y)\in\Lambda\times\Lambda$ and all $t>0$.  The
purpose of this section is to show that there is a representative of the
integral kernel, which is jointly continuous in $(t,x,y)$, $t>0$,
$x,y\in\Lambda$.  Moreover, this representative is given in terms of an
expectation with respect to the Brownian bridge.

To begin with, we collect some preparatory material concerning the Brownian
bridge. We define a continuous version $b$ of the {\bf Brownian bridge} from
$x=b(0)$ to $y=b(t)$ in terms of standard Brownian motion $w$ starting from
$x=w(0)$ by
\begin{equation}\label{6.4}
   b(s) \defeq
   \begin{cases} \Ds
      \frac st y + w(s) -
      (t-s)\int_0^s \frac{w(u)}{(t-u)^2}\,\d u
      &\Ds \text{\phantom{for }} 0\le s < t
   \\[-7pt] &  \text{for} \\[-7pt] \Ds
      y
      &\Ds \text{\phantom{for }} s=t
   \end{cases}.
\end{equation}
See \Cite[Section~5.6.B]{KaSh91} or \Cite[Exercise~IX.2.12]{ReYo94}. We note
that $b$ is adapted to the standard filtration \Cite[Definition~1.2.25]{KaSh91}
generated by $w$ and is a continuous semi-martingale
\Cite[Example~V.6.3]{Pro92}. The probability measure $\P_{0,x}^{t,y}$
associated with the Brownian bridge is the image of the Wiener measure $\P_{x}$
by the mapping $w\mapsto b$ as given by \xref{6.4}. The induced expectation is
written as $\E_{0,x}^{t,y}$.

For any stochastic process $s\mapsto Z(s)$, $0\le s<t$, adapted to the standard
filtration generated by $w$ with $\E_{x}[\,|Z(s)|\,]<\infty$ one has the
relation \Cite[Lemma~3.1]{Nak77}
\begin{equation}\label{6.5}
   \E_{0,x}^{t,y}[Z(s)]
   =
   \left(\frac{t}{t-s}\right)^{d/2}\:
   \e^{(x-y)^2/2t}\:
   \E_{x}\!\left[\,
      Z(s)\:
      \e^{-(w(s)-y)^2/2(t-s)}\:
   \,\right]
\end{equation}
for all $0\le s<t$ and all $x,y\in\rz^d$. This is a consequence of the
Cameron-Martin-Girsanov theorem \Cite[Theorem~3.5.1]{KaSh91}, since the
Brownian bridge is obtained from Brownian motion by adding a non-stationary
drift, see \xref{6.4}. We note that one may alternatively use the right-hand
side of \xref{6.5} to construct the Brownian-bridge measure, see
\Cite[Proposition~A.1]{Szn98}.

For a vector potential $A$ obeying
\begin{equation}\label{6.6}
   \P_{0,x}^{t,y}\!\left\{
      \int_0^t
         \left(A(b(s))\right)^2
      \,\d s <\infty
   \right\} = 1
\end{equation}
and
\begin{equation}\label{6.7}
   \P_{0,x}^{t,y}\!\left\{
      \left|\int_0^t
         A(b(s))\cdot
         \frac{y-b(s)}{t-s}
      \,\d s\right|
      < \infty
   \right\} = 1
\end{equation}
the stochastic line integral with respect to the Brownian bridge
\begin{equation}\label{6.8}\!\!\!\!\!\!\!\!
   s \mapsto
   \int_0^s A(b(u))\cdot\d b(u)
   \defeq
   \int_0^s A(b(u))\cdot\d w(u)
   +
   \int_0^s
         A(b(u))\cdot
         \frac{y-b(u)}{t-u}
   \,\d u
\end{equation}
is a well-defined stochastic process for $0\le s\le t$ having a continuous
version, confer \Cite[Definition~IV.2.9]{ReYo94}. According to Lemma~\ref{C-8},
our usual assumption $A\in\H_{\text{loc}}(\rz^d)$ suffices to ensure the
conditions \xref{6.6} and \xref{6.7}.

Now we are able to state the main result of this section:
\begin{structure}{Theorem}\label{6-1}\structindent\it
  Let $A\in\H_{\text{loc}}(\rz^d)$, $V\in\K_{\pm}(\rz^d)$, $1\le p\le q \le
  \infty$ and $\Lambda\subseteq\rz^d$ open. Recall the definitions \xref{2.33}
  and \xref{2.35} of $S_t$ and $\Xi_{\Lambda,t}$. Then
  \begin{equation}\label{6.9}
    k_t(x,y)
    \defeq
    \left(2\pi t\right)^{-d/2} \, \e^{-(x-y)^2/2t} \:
    \E_{0,x}^{t,y}\!\left[\,
      \e^{-S_t(A,V|b)}\,\Xi_{\Lambda,t}(b)\,
    \right]
  \end{equation}
  defines pointwise that integral kernel for
  \begin{equation}\label{6.10}
    \e^{-tH_\Lambda(A,V)}:\L^p(\Lambda)\to\L^q(\Lambda)
  \end{equation}
  which is jointly continuous in $(t,x,y)$, $t>0$, $x,y\in\Lambda$.
\end{structure}
\subsubsection*{Proof:}
Since the Brownian-bridge expectation in \xref{6.9} yields a regular version of
the conditional expectation in \xref{6.3}, $k_t$ as defined by \xref{6.9} is an
integral kernel for \xref{6.10}. Moreover, recalling our preparations, the
kernel $k_t$ is well defined for all $t>0$, $x,y\in\Lambda$. It remains to show
the claimed continuity of the function $k$ defined by \xref{6.9}.

Employing the time-reversal symmetry of the Brownian bridge we get from
\xref{6.9} the Hermiticity of the integral kernel, that is, $k_t(x,y)$ turns
into its complex conjugate upon interchanging $x$ and $y$. Therefore, it is
sufficient to ensure
\begin{equation}\label{6.11}
  \lim_{\rho\downarrow 0} \;
  \sup_{\tau_1\le t\le t'\le\tau_2, |t-t'|<\rho}\;
  \sup_{y,y'\in K, |y-y'|<\rho}\;
  \sup_{x\in K}\;
  | k_t(x,y) - k_{t'}(x,y') | = 0
\end{equation}
for all compact $K\subset\Lambda$ and all $0<\tau_1\le\tau_2<\infty$.  Note
that we have assumed $t\le t'$.

For any $0<s<\tau_1$ we can rewrite the difference in \xref{6.11} according to
\begin{equation}\label{6.12}
  k_t(x,y) - k_{t'}(x,y') =
  \Upsilon(t',t'-t+s,x,y') - \Upsilon(t,s,x,y)
  + \Gamma(t,t',s,x,y,y'),
\end{equation}
where we have introduced the abbreviations
\begin{equation}\label{6.13}
  \begin{split}\Ds
    \Upsilon(t,s,x,y) \defeq
  & \Ds
    \left(2\pi t\right)^{-d/2} \, \e^{-(x-y)^2/2t} \:
  \\ \Ds  &\Ds \hbox{} \times
    \E_{0,x}^{t,y}\!\left[\,
      \e^{-S_t(A,V|b)}\,\Xi_{\Lambda,t}(b)
      - \e^{-S_{t-s}(A,V|b)}\,\Xi_{\Lambda,t-s}(b)\,
    \right]
  \end{split}
\end{equation}
and
\begin{equation}\label{6.14}
  \begin{split}\Ds
    \Gamma(t,t',s,x,y,y') \defeq
  & \Ds
    \left(2\pi t'\right)^{-d/2} \, \e^{-(x-y')^2/2t'} \:
    \E_{0,x}^{t',y'}\!\left[\,
      \e^{-S_{t-s}(A,V|b)}\,\Xi_{\Lambda,t-s}(b)\,
    \right]
  \\ \Ds  &\Ds \hbox{} -
    \left(2\pi t\right)^{-d/2} \, \e^{-(x-y)^2/2t} \:
    \E_{0,x}^{t,y}\!\left[\,
      \e^{-S_{t-s}(A,V|b)}\,\Xi_{\Lambda,t-s}(b)\,
    \right].
  \end{split}
\end{equation}
Therefore, it is enough to show
\begin{equation}\label{6.15}
  \lim_{s\downarrow 0} \;
  \sup_{\tau_1\le t\le\tau_2}\;
  \sup_{x,y\in K}\;
  | \Upsilon(t,s,x,y) | = 0
\end{equation}
and
\begin{equation}\label{6.16}
  \lim_{\rho\downarrow 0} \;
  \sup_{\tau_1\le t\le t'\le\tau_2, |t-t'|<\rho}\;
  \sup_{y,y'\in \rz^d, |y-y'|<\rho}\;
  \sup_{x\in\rz^d}\;
  | \Gamma(t,t',s,x,y,y') | = 0
\end{equation}
for all compact $K\subset\Lambda$ and all $0<s<\tau_1\le\tau_2<\infty$.  The
assertion \xref{6.16} is actually stronger than needed for the present purpose,
but will turn out to be useful in the sequel.

From the triangle inequality we get
\begin{equation}\label{6.17}
  \begin{split}\Ds
    \Upsilon(t,s,x,y) \le
  & \Ds
    \left(2\pi t\right)^{-d/2} \, \e^{-(x-y)^2/2t} \:
  \\ \Ds  &\Ds \hbox{} \times \Bigl(\,
    \E_{0,x}^{t,y}\!\left[\,\left|
      \e^{-S_t(A,V|b)} - \e^{-S_{t-s}(A,V|b)}
    \right| \right]
  \\ \Ds  &\Ds \hbox{} \phantom{\times \biggl(\,} +
    \E_{0,x}^{t,y}\!\left[\,
      \e^{-S_{t-s}(0,V|b)}\,
      \left| \, \Xi_{\Lambda,t}(b) - \Xi_{\Lambda,t-s}(b)\,\right|\,
    \right]
  \Bigr).
  \end{split}
\end{equation}
Now we make use of the Cauchy-Schwarz inequality, the elementary estimate
\begin{equation}\label{6.18}
  S_{t-s}(0,2V|b) \ge S_t(0,-2V^-|b)
\end{equation}
and the time-reversal symmetry of the Brownian bridge. We thus achieve
\begin{equation}\label{6.19}
  \begin{split}\Ds
    \Upsilon(t,s,x,y) \le \hbox{}
  & \Ds
    N^{1/2}\,\left(2\pi t\right)^{-d/4} \, \e^{-(x-y)^2/4t}
  \\ \Ds  &\Ds \hbox{} \times 
    \biggl(\,
      \left(\E_{0,y}^{t,x}\!\left[\,\left|
        1 - \e^{-S_s(A,V|b)}
      \right|^2 \right]\right)^{1/2}
      + 
      \bigl(\E_{0,y}^{t,x}\!\left[\,
        1 - \Xi_{\Lambda,s}(b)
      \right]\bigr)^{1/2}
    \biggr)
  \\ \Ds \hbox{} \le \hbox{} &\Ds 
    N^{1/2}\,\left(2\pi (t-s)\right)^{-d/4} 
  \\ \Ds  &\Ds \hbox{} \times 
    \biggl(\,
      \left(\E_y\!\left[\,\left|
        1 - \e^{-S_s(A,V|w)}
      \right|^2 \right]\right)^{1/2}
      + 
      \bigl(\E_y\!\left[\,
        1 - \Xi_{\Lambda,s}(w)
      \right]\bigr)^{1/2}
    \biggr).
  \end{split}
\end{equation}
Here we have used \xref{6.5} for the second step and have set
\begin{equation}\label{6.20}
  N \defeq
  \sup_{\tau_1\le t\le\tau_2}\;
  \sup_{x,y\in\rz^d}\;
  \left(2\pi t\right)^{-d/2} \, \e^{-(x-y)^2/2t} \:
  \E_{0,x}^{t,y}\!\left[\,
    \e^{-S_t(0,-2V^-|b)}\,
  \right].
\end{equation}
Employing the Cauchy-Schwarz inequality, the time-reversal symmetry and
\xref{6.5} again, we get
\begin{equation}\label{6.21}
  N \le
  (\pi\tau_1)^{-d/2}
  \sup_{x\in\rz^d}\;
  \E_x\!\left[\,
    \e^{-S_{\tau_2/2}(0,-4V^-|w)}\,
  \right],
\end{equation}
which shows by virtue of Lemma~\ref{C-2} that $N$ is finite. Note that it is
essential to take the supremum in \xref{6.20} and not only the essential
supremum as, for example, in \xref{6.2}. Having shown that $N$ is finite,
\xref{6.15} follows from \xref{6.19} with the help of \xref{3.11} and
Lemma~\ref{C-5}.

The remaining task is to prove \xref{6.16}. In a first step we use \xref{6.5}
to rewrite $\Gamma$ as given by \xref{6.14} in terms of an expectation with
respect to Brownian motion. Then the Cauchy-Schwarz inequality and \xref{6.18}
lead to
\begin{equation}\label{6.22}
  \begin{split}\Ds
    \left|\,\Gamma(t,t',s,x,y,y')\, \right|^2 \le\hbox{}
  & \Ds
    (2\pi)^{-d}\,M \: \E_x\!\biggl[\,\Bigl(\,
      (t'-t+s)^{-d/2} \, \e^{-(w(t-s)-y')^2/2(t'-t+s)} 
  \\ \Ds  &\Ds \phantom{(2\pi)^{-d}\,M \: \E_x\!\biggl[\,\Bigl(\,}
      -
      s^{-d/2} \, \e^{-(w(t-s)-y)^2/2s} 
    \Bigr)^{\! 2}\biggr],
  \end{split}
\end{equation}
where
\begin{equation}\label{6.23}
  M \defeq
  \sup_{x\in\rz^d}\;
  \E_x\!\left[\,
    \e^{-S_{\tau_2}(0,-2V^-|w)}\,
  \right]
\end{equation}
is finite due to Lemma~\ref{C-2}. The expectation on the right-hand side of
\xref{6.22} can be calculated explicitly. The result implies
\begin{equation}\label{6.24}
  \begin{split}\Ds & \Ds
    \left|\,\Gamma(t,t',s,x,y,y')\, \right|^2 
  \\ \Ds \hbox{}\le\hbox{} &\Ds (2\pi)^{-d}\,M\biggl[\,
    \left({t'}^2-(t-s)^2\right)^{-d/2}
    \exp\!\left\{
      - \frac{(x-y')^2}{t'+t-s}
    \right\}
  \\ \Ds  &\Ds \phantom{(2\pi)^{-d}\,M\biggl[\,} +
    s^{-d/2}\,(2t-s)^{-d/2}
    \exp\!\left\{
      - \frac{(x-y)^2}{2t-s}
    \right\}
  \\ \Ds  &\Ds \phantom{(2\pi)^{-d}\,M\biggl[\,} -
    2\left(t\,t'-(t-s)^2\right)^{-d/2}
  \\ \Ds &\Ds \phantom{(2\pi)^{-d}\,M\biggl[\,-}\times
    \exp\!\left\{
      - 
      \frac{s(x-y')^2 + (t-s)(y-y')^2 + (t'-t+s)(x-y)^2}{2(t\,t'-(t-s)^2)}
    \right\}
    \,\biggr].
  \end{split}
\end{equation}
Now we use the three elementary inequalities
\begin{equation}\label{6.25}\!\!\!
  \frac{1}{{t'}^2-(t-s)^2} \le \frac{1}{s(2t-s)}, \quad
  \frac{1}{2t-s} \ge   \frac{1}{t'+t-s}, \quad 
  \frac{1}{t\,t'-(t-s)^2} \le   \frac{1}{s(t'+t-s)},
\end{equation}
valid for all $0<s<t\le t'<\infty$, and get
\begin{equation}\label{6.26}
  \begin{split}\Ds & \Ds
    \left|\,\Gamma(t,t',s,x,y,y')\, \right|^2 
  \\ \Ds \hbox{}\le\hbox{} &\Ds 
     2(2\pi)^{-d}\,M\,s^{-d/2}\,(2t-s)^{-d/2}
    \exp\!\left\{
      - \frac{1}{2} \, \frac{(x-y')^2+(x-y)^2}{t'+t-s}
    \right\}
  \\ \Ds  &\Ds \hbox{} \times \biggl[\,
    \cosh\!\left\{
      \frac{(y-y')\cdot(y+y'-2x)}{2(t+t'-s)}
    \right\}
  \\ \Ds  &\Ds \phantom{\hbox{} \times \biggl[\,} -
    \left( 1+\frac{t(t'-t)}{s(2t-s)} \right)^{-d/2}
    \exp\!\left\{
      - \frac{(t-s)(y-y')^2+(t'-t)(x-y)^2}{2s(t+t'-s)}
    \right\}
    \,\biggr].
  \end{split}
\end{equation}
In a final step we further estimate the right-hand side with the help of the
elementary inequalities
\begin{equation}\label{6.27}
  \begin{split}\Ds & \Ds
    \cosh(a)\le\frac{1}{2}a^2\cosh(a)+1,\quad a\in\rz, \quad
  \\ \Ds & \Ds
    \left(1+b\right)^{-d/2}\ge 1-\frac{d}{2}b, \quad b>0,\qquad
    \e^{-c}\ge 1-c,\quad c\in\rz,
  \end{split}
\end{equation}
the first one of which may readily be derived from the Taylor expansion
\Cite[Equation~4.5.63]{AbSt72} of the hyperbolic cosine. The assertion
\xref{6.16} then follows from the resulting bound by inspection.\halmos
\begin{structure}{Remarks}\label{6-2}\structindent
\structitem\label{6-2-i} Theorem~\ref{6-1} generalizes
  \Cite[Theorem~B.7.1.(a$^{\prime\prime\prime}$)]{Sim82} to non-zero
  $A\in\H_{\text{loc}}(\rz^d)$ and $\Lambda\subseteq\rz^d$. The proof given
  there relies on the local-norm-continuity of the semigroup in the scalar
  potential and the approximability Proposition~\ref{2-3} (see
  \Cite[Theorem~B.10.2]{Sim82}, but beware of a circularity in the proofs of
  \Cite[Proposition~B.6.7, Theorem~B.7.1.(a$^\prime$), Lemma~B.7.4]{Sim82}). By
  using Theorem~\ref{5-1} local-norm-continuity covers also the case $A\not=0$.
 
  Results similar to Theorem~\ref{6-1} for $A=0$ and $\Lambda\subseteq\rz^d$
  are contained in \Cite[Theorem~3.17]{ChZh95} and
  \Cite[Proposition~1.3.5]{Szn98}.
  
  Like the last reference our approach has the advantage of singling out the
  continuous representative of the integral kernel as a Brownian-bridge
  expectation, which is useful to know for further investigations and
  applications.

  The rather general Theorem~14.5 in \cite{Sim79a} does not cover the
  above Theorem~\ref{6-1}. A discriminating example corresponds to the 
  hydrogen atom: $\Lambda=\rz^3$, $A=0$, $V(x)=-|x|^{-1}$. In the
  notation of \cite{Sim79a} this means: $\nu=3$, $f=0$, $F(a)=\e^{ia}$ 
  for $a\in\rz$, $g(\text{\boldmath$\omega$}) = 
  \exp\!\left\{\int_0^t|\text{\boldmath$\omega$}(s)|^{-1}\d s\right\}$
  for $t>0$. Theorem~14.5 in \cite{Sim79a} is not applicable, because
  the Wiener-essential supremum $\Vert g\Vert_\infty$ of $g$ is infinite.
\structitem\label{6-2-ii} The continuity of the integral kernel of the
  semigroup implies the continuity of the integral kernels for various
  functions of the Schr\"odinger operator $H_\Lambda(A,V)$. This can be shown
  along the same lines of reasoning as in the proof of
  \Cite[Theorem~B.7.1]{Sim82}, since the arguments given there do not use the
  fact that $H_{\rz^d}(0,V)$ is a particular Schr\"odinger operator, but merely
  the continuity of $k_t(x,y)$ and the bound \xref{2.40}.  For example, each
  spectral projection
   \begin{equation}\label{6.28}
      \chi_{_{\scriptstyle\Omega}}(H_\Lambda(A,V)) : 
      \L^2(\Lambda) \to  \L^2(\Lambda)
   \end{equation}
   associated with a bounded Borel subset $\Omega\subset\rz$ is 
   an integral operator with jointly continuous kernel. 
\structitem\label{6-2-iii} Since $k_t(x,y)$ is jointly continuous in $(x,y)$
  for $t>0$, the trace of the non-negative operator $\e^{-tH_\Lambda(A,V)} :\:
  \L^2(\Lambda) \rightarrow \L^2(\Lambda)$ can be represented as
  \begin{equation}\label{6.29}
    {\rm tr}\: \e^{-tH_\Lambda(A,V)}
    =
    \int_\Lambda k_t(x,x)\,\d x,
  \end{equation}
  whenever one of the two sides of this equation is finite.
\structitem\label{6-2-iv} Due to the continuity of the integral kernel, the
  diamagnetic inequality \xref{2.38} takes the form
  \begin{equation}\label{6.30}
    \left| k_t(x,y) \right|
    \le
    \left. k_t(x,y) \right|_{A=0}
  \end{equation}
  and is valid pointwise for all $x,y\in\Lambda$, $t>0$. Various upper bounds
  on the right-hand side of \xref{6.30} are available in the literature, see,
  for example, the estimate \Cite[Theorem~B.7.1.(a$^\prime$)]{Sim82}. Using
  this estimate, Theorem~\ref{4-1} can be proven alternatively with the help of
  the continuity of the integral kernel and the dominated-convergence theorem.
  We have given a different proof of Theorem~\ref{4-1}, mainly because we think
  it is nice to see how Carmona's elegant argument for $A=0$ can be extended to
  $A\not=0$.
\end{structure}
\begin{structure}{Theorem}\label{6-3}\structindent\it
  Let $A\in\H_{\text{loc}}(\rz^d)$, $V\in\K_{\pm}(\rz^d)$ and
  $\Lambda\subseteq\rz^d$ regular. Then $k_t(x,y)$, as defined by \xref{6.9},
  is jointly continuous in $(t,x,y)$, $t>0$, $x,y\in\overline{\Lambda}$ and
  vanishes if $x\in\partial\Lambda$ or $y\in\partial\Lambda$.
\end{structure}
\subsubsection*{Proof:}
From \xref{6.9} we derive with the help of the Cauchy-Schwarz inequality and
$\Xi_{\Lambda,t}(b)\le\Xi_{\Lambda,t/2}(b)$ the estimate
\begin{equation}\label{6.31}
  \left| k_t(x,y) \right| \le
  \left(2\pi t\right)^{-d/4} \, N^{1/2} \, \e^{-(x-y)^2/4t} \:
  \left( \E_{0,x}^{t,y}\!\left[\,
     \Xi_{\Lambda,t/2}(b)\,
  \right]\right)^{1/2},
\end{equation}
where $0\le N<\infty$ is given by \xref{6.20}. With the help of \xref{6.5} we
achieve
\begin{equation}\label{6.32}
  \left| k_t(x,y) \right| \le
  \left(\pi t\right)^{-d/4} \, N^{1/2} \, 
  \left( \E_x\!\left[\,
     \Xi_{\Lambda,t/2}(b)\,
  \right]\right)^{1/2}.
\end{equation}
According to Lemma~\ref{C-7} the right-hand side vanishes as $x$ approaches any
given point on the boundary $\partial\Lambda$. Due to the Hermiticity of $k_t$
and Theorem~\ref{6-1} the assertion follows.\halmos
\begin{structure}{Remark}\label{6-4}\structindent
  Theorem~\ref{6-3} is a partial generalization of \Cite[Theorem~3.17]{ChZh95}
  to $A\neq 0$.
\end{structure}
We finally want to show that the same strategy as used in Section~\ref{4}
allows to get uniform continuity for global regularity conditions.
\begin{structure}{Theorem}\label{6-5}\structindent\it
  Let $A\in\H(\rz^d)$, $V\in\K(\rz^d)$, $0<\tau_1<\tau_2<\infty$ and
  $\Lambda\subseteq\rz^d$ uniformly regular. Then the function
  \begin{equation}\label{6.33}
    [\tau_1,\tau_2]\times\overline{\Lambda}\times\overline{\Lambda}
    \to \cz, \quad (t,x,y) \mapsto k_t(x,y)
  \end{equation}
  is uniformly continuous.
\end{structure}
\subsubsection*{Proof:}
Analogously to the reasoning at the beginning of the proof of
Theorem~\ref{6-1}, it is sufficient to ensure
\begin{equation}\label{6.34}
  \lim_{\rho\downarrow 0} \;
  \sup_{\tau_1\le t\le t'\le\tau_2, |t-t'|<\rho}\;
  \sup_{y,y'\in \Lambda_r, |y-y'|<\rho}\;
  \sup_{x\in \Lambda_r}\;
  | k_t(x,y) - k_{t'}(x,y') | = 0
\end{equation}
in order to get uniform continuity on
$[\tau_1,\tau_2]\times\Lambda_r\times\Lambda_r$, $r>0$, where $\Lambda_r$ is
given by \xref{3.12}. Due to \xref{6.12} and \xref{6.16} we only have to check
that
\begin{equation}\label{6.35}
  \lim_{s\downarrow 0} \;
  \sup_{\tau_1\le t\le\tau_2}\;
  \sup_{x,y\in\Lambda_r}\;
  | \Upsilon(t,s,x,y) | = 0,
\end{equation}
where $\Upsilon$ is given by \xref{6.13}. With the help of the estimate
\xref{6.19} this follows from Lemma~\ref{C-3} and \xref{3.11}.

On the other hand, Lemma~\ref{C-7} and \xref{6.32} imply
\begin{equation}\label{6.36}
  \lim_{r\downarrow 0} \;
  \sup_{\tau_1\le t\le\tau_2}\;
  \sup_{x\in\Lambda_r}\;
  \sup_{y\in\Lambda\backslash\Lambda_r}\;
  | k_t(x,y) | = 0,
\end{equation}
which suffices to extend the domain of uniform continuity from
$[\tau_1,\tau_2]\times\Lambda_r\times\Lambda_r$, $r>0$, to
$[\tau_1,\tau_2]\times\overline{\Lambda}\times\overline{\Lambda}$.\halmos
\begin{appendix}
\Section{Approximability of Kato-type potentials by smooth functions}
\label{A}
We want to prove Propositions~\ref{2-3} and \ref{2-6}. The idea of the proof
has been given already in Section~\ref{2}. It is similar to the one underlying
the standard proof of the fact that $\C_0^\infty(\rz^d)$ is dense in
$\L^p(\rz^d)$ for $1\le p<\infty$.

To begin with, we choose $\delta_1\in\C_0^\infty(\rz^d)$ such that $\delta_1\ge
0$, $\delta_1(x)=0$ for $|x|>1$ and $\|\delta_1\|_1=1$. Moreover, we set
$\delta_r(x)\defeq r^{-d}\delta_1(x/r)$ for $0<r\le 1$ and
$\Theta_R\defeq\delta_1\ast\chi_{_{\scriptstyle B_R}}$ for $R>1$, where
\begin{equation}\label{A.1}
   (h\ast F)(x) \defeq \int
   h(x-y)\:F(y)\:\d y
\end{equation}
denotes the convolution of a function $F:\rz^d\to\rz^\nu$, $\nu\in\nz$, with a
function $h:\rz^d\to\rz$. We note that $\Theta_R\in\C_0^\infty(\rz^d)$ and
\begin{equation}\label{A.2}
   \chi_{_{\scriptstyle B_{R-1}}}
   \le \Theta_R \le
   \chi_{_{\scriptstyle B_{R+1}}}.
\end{equation}
Now we are ready to formulate the following preparatory result.
\begin{structure}{Lemma}\label{A-1}\structindent\it
   Let $F:\rz^d\to\rz^\nu$,
   $|F|^p\in\K_{\text{loc}}(\rz^d)$, $1\le p<\infty$.
   Then
   \begin{gather}\label{A.3}
      \delta_r\ast\Theta_R F\in\left(\C_0^\infty(\rz^d)\right)^\nu,
   \\ \label{A.4}
      \left\|
         g_\rho\ast\left|\delta_r\ast\Theta_R F\right|^p
      \right\|_\infty
      \le
      \left\|
         g_\rho\ast\chi_{_{\scriptstyle B_{R+1}}}|F|^p
      \right\|_\infty
   \end{gather}
   and
   \begin{equation}\label{A.5}
      \lim_{r\downarrow 0}
      \left\|
         \left|
            \Theta_R F -
            \delta_r\ast\Theta_R F
         \right|^p
      \right\|_{\K(\rz^d)}
      =0
   \end{equation}
   for all $0<r\le 1<R<\infty$ and $0<\rho\le 1$.
\end{structure}
\subsubsection*{Proof:}
\xref{A.3} holds, since $\Theta_RF\in\Ds\left(\L^1(\rz^d)\right)^\nu$ by
\xref{2.12}.

From the Jensen inequality and \xref{A.2} we have
\begin{equation}\label{A.6}
   \left|\delta_r\ast\Theta_RF\right|^p
   \le
   \delta_r\ast\chi_{_{\scriptstyle B_{R+1}}}|F|^p.
\end{equation}
Furthermore, by the commutativity and the associativity of the convolution and
the monotonicity of integration we get
\begin{equation}\label{A.7}
   g_\rho\ast\delta_r\ast\chi_{_{\scriptstyle B_{R+1}}}|F|^p
   \le
   \left\|\delta_r\right\|_1\:
   \left\|
      g_\rho\ast\chi_{_{\scriptstyle B_{R+1}}}|F|^p
   \right\|_\infty.
\end{equation}
Now \xref{A.4} follows from \xref{A.6}, \xref{A.7} and $\|\delta_r\|_1=1$.

Since $\|f\|_{\K(\rz^d)}=\|g_1\ast|f|\|_\infty$, we observe for
$f\in\K(\rz^d)\cap\L^1(\rz^d)$ that
\begin{equation}\label{A.8}
   \|f\|_{\K(\rz^d)}
   \le
   \left\|g_\rho\ast|f|\right\|_\infty
   + G_\rho\,\left\|f\right\|_1,
\end{equation}
where
\begin{equation}\label{A.9}
   G_\rho\defeq\sup_{\rho\le|x|\le 1}g_1(x).
\end{equation}
On the other hand, the inequality $|u-v|^p\le 2^p\left(|u|^p+|v|^p\right)$ for
$u,v\in\rz^\nu$ in combination with \xref{A.2} and \xref{A.6} yields
\begin{equation}\label{A.10}
   \left|
      \Theta_RF -
      \delta_r\ast\Theta_RF
   \right|^p
   \le
   2^p\,\left(
      \chi_{_{\scriptstyle B_{R+1}}}|F|^p +
      \delta_r\ast\chi_{_{\scriptstyle B_{R+1}}}|F|^p
   \right).
\end{equation}
With the help of \xref{A.8}, \xref{A.10} and \xref{A.4} we establish
\begin{equation}\label{A.11}
   \begin{split} \Ds & \Ds
      \left\|
         \left|
            \Theta_RF -
            \delta_r\ast\Theta_RF
         \right|^p
      \right\|_{\K(\rz^d)}
   \\ \Ds \le \; & \Ds
      2^{p+1}
      \left\|
         g_\rho\ast\chi_{_{\scriptstyle B_{R+1}}}|F|^p
      \right\|_\infty
      +
      G_\rho\,\left\|
         \left|
            \Theta_RF -
            \delta_r\ast\Theta_RF
         \right|^p
      \right\|_1.
   \end{split}
\end{equation}
Since $\Theta_RF\in\Ds\left(\L^p(\rz^d)\right)^\nu$, the second term on the
right-hand side tends to zero as $r\downarrow 0$, see, for example,
\Cite[Lemma~1.3.6/2]{Tri92}. Moreover, the remaining first term vanishes as
$\rho\downarrow 0$ because $\chi_{_{\scriptstyle B_{R+1}}}|F|^p\in\K(\rz^d)$.
This proves \xref{A.5}.\halmos
\subsubsection*{Proof of Proposition~\ref{2-3}:} 
By the triangle inequality
\begin{equation}\label{A.12}
   \left|
      V - \delta_r\ast\Theta_RV
   \right|
   \,\chi_{_{\scriptstyle K}}
   \le
   V\,(1-\Theta_R)\,\chi_{_{\scriptstyle K}}
   +
   \left|
      \Theta_RV - \delta_r\ast\Theta_RV
   \right|
\end{equation}
and the fact that
\begin{equation}\label{A.13}
   (1-\Theta_R)\,\chi_{_{\scriptstyle K}} =0
   \quad\text{for}\quad
   R>1+\sup_{x\in K}|x|
\end{equation}
the approximability \xref{2.15} follows from \xref{A.3} and \xref{A.5}.  The
inequality \xref{2.16} follows from \xref{A.4} for the choice $p=1$ and
$F=V^-$.\halmos
\subsubsection*{Proof of Proposition~\ref{2-6}:}
We start from the pair of elementary inequalities
\begin{equation}\label{A.14}\!\!\!\!\!\!\!\!
   \left(
      A - \delta_r\ast\Theta_RA
   \right)^2
   \,\chi_{_{\scriptstyle K}}
   \le
   2A^2\,(1-\Theta_R)\,\chi_{_{\scriptstyle K}}
   +
   2\left(
      \Theta_RA - \delta_r\ast\Theta_RA
   \right)^2
\end{equation}
and
\begin{equation}\label{A.15}
   \begin{split}\Ds & \Ds
      \left|
         \nabla\cdot A - \nabla\cdot(\delta_r\ast\Theta_RA)
      \right|
      \,\chi_{_{\scriptstyle K}}
   \\ \Ds \le \; & \Ds
      |\nabla\cdot A|\,(1-\Theta_R)\,\chi_{_{\scriptstyle K}}
      +
      \left|
         \delta_r\ast(A\cdot\nabla\Theta_R)
      \right|\,\chi_{_{\scriptstyle K}}
   \\ \Ds &\Ds\hbox{}
      +
      \left|
         \Theta_R\nabla\cdot A -
         \delta_r\ast(\Theta_R\nabla\cdot A)
      \right|.
   \end{split}
\end{equation}
The second term on the right-hand side of \xref{A.15} can be estimated further
with the help of \xref{A.2} according to
\begin{equation}\label{A.16}\!\!\!\!\!\!\!\!\!\!\!\!\!\!\!\!
   \left|
      \delta_r\ast(A\cdot\nabla\Theta_R)
   \right|\,\chi_{_{\scriptstyle K}}
   \le
   \chi_{_{\scriptstyle K}}\,
   \chi_{_{\scriptstyle B_{R+2}}}\,
   \left(1-\chi_{_{\scriptstyle B_{R-2}}}\right)\,
   \left(\delta_r\ast|A|\right)\,
   \sup_{x\in\rz^d}\left|(\nabla\Theta_R)(x)\right| \!\!\!\!\!\!\!\!
\end{equation}
and, hence, vanishes for $R>2+\sup_{x\in K}|x|$. In consequence, the
approximability \xref{2.19}, \xref{2.20} follows from \xref{A.3} with the help
of \xref{A.13} and \xref{A.5}.\halmos
\Section{Proof of the Feynman-Kac-It\^o formula}
\label{B}
The purpose of this appendix is to prove Proposition~\ref{2-9}. The strategy of
the proof has already been sketched in Section~\ref{2}.  Here we only recall
that we take \xref{2.34} in the case $\Lambda=\rz^d$,
$A\in\H_{\text{loc}}(\rz^d)$ and $V\in\K_{\pm}(\rz^d)$ for granted.

Let $\{K_l\}_{l\in\nz}$ be a strictly increasing sequence of compact subsets of
$\Lambda$ that exhausts $\Lambda$. That is, each $K_l$ is contained in the
interior of $K_{l+1}$ and $\bigcup_{l\in\nz}K_l=\Lambda$. Furthermore let
$\{\theta_l\}_{l\in\nz}\subset\C_0^\infty(\rz^d)$ be such that $\theta_l(x)=1$
for $x\in K_l$, $\theta_l(x)=0$ for $x\not\in K_{l+1}$, and $0\le\theta_l(x)\le
1$ for all $x\in\rz^d$. Define $U_\infty^\Lambda:\rz^d\to\rz$ by
\begin{equation}\label{B.1}
   U_\infty^\Lambda(x) \defeq
  \begin{cases}\Ds
     \sum_{l\in\nz} \left|(\nabla\theta_l)(x)\right|^2
     + \left(\inf\{|x-y|:y\not\in\Lambda\}\right)^{-3}    
            & \Ds \text{for $x\in\Lambda$} \\ \Ds
     + \infty & \Ds \text{otherwise}
  \end{cases}
\end{equation}
and $U_n^\Lambda:\rz^d\to\rz$ for $n\in\nz$ by
\begin{equation}\label{B.2}
   U_n^\Lambda(x) \defeq \inf\{U_\infty^\Lambda(x),n\}.
\end{equation}

The sesquilinear form $h_{\rz^d}^{A,\,\mu U_\infty^\Lambda+V}$,
$A\in\H_{\text{loc}}(\rz^d)$, $V\in\L^\infty(\rz^d)$, $\mu>0$, defined through
\xref{2.22}, with domain $\Q\!\left(h_{\rz^d}^{A,\,\mu
    U_\infty^\Lambda+V}\right)= \Q\!\left(h_{\rz^d}^{A,
    U_\infty^\Lambda}\right)$ given by \xref{2.23} is closed but not densely
defined. In fact $\C_0^\infty(\Lambda)\subset
\Q\!\left(h_{\rz^d}^{A,U_\infty^\Lambda}\right) \subset\L^2(\Lambda)$ by the
extension convention of Section~\ref{2} and the form is densely defined on
$\L^2(\Lambda)$. It gives rise to a self-adjoint non-negative operator on
$\L^2(\Lambda)$ which we denote by $H_{\rz^d}(A,\mu U_\infty^\Lambda+V)$. For a
function
\begin{equation}\label{B.3}
   f:\rz \to \rz,\quad
   \text{$f$ continuous},\quad
   \lim_{x\to +\infty}f(x)=0
\end{equation}
we define for $\psi\in\L^2(\rz^d)$
\begin{equation}\label{B.4}
   f\!\left(H_{\rz^d}(A,\mu U_\infty^\Lambda+V)\right)\psi \defeq 
   f\!\left(H_{\rz^d}(A,\mu U_\infty^\Lambda+V)\right)
   (\chi_{_{\scriptstyle\Lambda}}\,\psi) .
\end{equation}
Hence $f\!\left(H_{\rz^d}(A,\mu U_\infty^\Lambda+V)\right)$ naturally extends
to a bounded self-adjoint operator on $\L^2(\rz^d)$.

Our plan for the proof of Proposition~\ref{2-9} is summarized in the following
six lemmas.
\begin{structure}{Lemma}\label{B-1}\structindent\it
   Let $A\in\H_{\text{loc}}(\rz^d)$, $V\in\L^\infty(\rz^d)$, $V\ge 0$,
   $\psi\in\L^2(\rz^d)$, $\mu>0$, $t\ge 0$, $\Lambda\subseteq\rz^d$ open. Then
   \begin{equation}\label{B.5}
      \lim_{n\to\infty} 
      \left\|\,
         \left(
            \e^{-tH_{\rz^d}(A,\,\mu U_n^\Lambda+V)} -
            \e^{-tH_{\rz^d}(A,\,\mu U_\infty^\Lambda+V)}
         \right)\psi
      \,\right\|_{2}
      = 0.
   \end{equation}
\end{structure}
Lemma~\ref{B-1} serves for the proof of the Feynman-Kac-It\^o formula for
$\e^{-tH_{\rz^d}(A,\,\mu U_\infty^\Lambda+V)}$.
\begin{structure}{Lemma}\label{B-2}\structindent\it
   Let $A\in\H_{\text{loc}}(\rz^d)$, $V\in\L^\infty(\rz^d)$, $V\ge 0$,
   $\psi\in\L^2(\rz^d)$, $\mu>0$, $t\ge 0$, $\Lambda\subseteq\rz^d$ open. Then
   \begin{equation}\label{B.6}
      \left(\e^{-tH_{\rz^d}(A,\,\mu U_\infty^\Lambda+V)}\psi\right)(x) =
      \E_{x}\!\left[
          \e^{-S_t(A,\,\mu U_\infty^\Lambda+V|w)}\,
          \Xi_{\Lambda,t}(w)\,\psi(w(t))
      \right]
   \end{equation}
   for almost all $x\in\rz^d$.
\end{structure}
The purpose of Lemma~\ref{B-2} is twofold. First, it immediately implies the
inequality
\begin{equation}\label{B.7}
   |\e^{-tH_{\rz^d}(A,\,\mu U_\infty^\Lambda+V)}\psi|
   \le\e^{-tH_{\rz^d}(0,0)}|\psi|,\quad \psi\in\L^2(\Lambda),\; t\ge 0,
\end{equation}
which is crucial in the proof of the following lemma.
\begin{structure}{Lemma}\label{B-3}\structindent\it
  Let $A\in\H_{\text{loc}}(\rz^d)$, $\Lambda\subseteq\rz^d$ open. Then
  the completion of the form domain
  $\Q\!\left(h_{\rz^d}^{A,U_\infty^\Lambda}\right)$ with respect to
  the form-norm $\|\bullet\|_{h_{\rz^d}^{A,0}}$ is
  $\Q\!\left(h_\Lambda^{A,0}\right)$.
\end{structure}
This rather technical lemma singles out the difficult part of the proof of the
next approximation.
\begin{structure}{Lemma}\label{B-4}\structindent\it
  Let $A\in\H_{\text{loc}}(\rz^d)$, $V\in\L^\infty(\rz^d)$, $V\ge 0$,
  $\psi\in\L^2(\rz^d)$, $\Lambda\subseteq\rz^d$ open, $t\ge 0$. Then
   \begin{equation}\label{B.8}
      \lim_{\mu\downarrow 0} 
      \left\|\,
         \left(
            \e^{-tH_{\rz^d}(A,\,\mu U_\infty^\Lambda+V)} -
            \e^{-tH_\Lambda(A,V)} 
         \right)\psi
      \,\right\|_{2}
      = 0.
   \end{equation}
\end{structure}
The second purpose of Lemma~\ref{B-2} is, of course, to be used together with
\ref{B-4} to get the next variant of the Feynman-Kac-It\^o formula. Assume
$A\in\H_{\text{loc}}(\rz^d)$, $V\in\L^\infty(\rz^d)$, $V\ge 0$. Then
Lemma~\ref{B-4} implies that the left-hand side of \xref{B.6} tends to the
left-hand side of \xref{2.34} for almost all $x\in\Lambda$ as $\mu\downarrow 0$
at least for a suitably chosen subsequence. Furthermore, the right-hand side of
\xref{B.6} converges to the right-hand side of \xref{2.34} for almost all
$x\in\Lambda$ as $\mu\downarrow 0$ by the dominated-convergence theorem. One
may relax the assumption $V\in\L^\infty(\rz^d)$, $V\ge 0$ to
$V\in\L^\infty(\rz^d)$ by multiplying both sides of \xref{2.34} with
$\e^{t\inf_{x\in\rz^d}V(x)}$. We summarize these findings in the following
lemma.
\begin{structure}{Lemma}\label{B-5}\structindent\it
  Let $A\in\H_{\text{loc}}(\rz^d)$, $V\in\L^\infty(\rz^d)$,
  $\psi\in\L^2(\Lambda)$, $\Lambda\subseteq\rz^d$ open, $t\ge 0$. Then
  \xref{2.34} holds for almost all $x\in\Lambda$.
\end{structure}
With this statement at hand, the diamagnetic inequality \xref{2.25} follows
from the triangle inequality. This is just-in-time because we used \xref{2.25}
in the case $\Lambda\subset\rz^d$, $\Lambda\not=\rz^d$ for the construction of
$H_\Lambda(A,V)$, $A\in\H_{\text{loc}}(\rz^d)$, $V\in\K_{\pm}(\rz^d)$, in that
case.

Now we are ready to formulate the last approximation tool for the proof of
Proposition~\ref{2-9}.
\begin{structure}{Lemma}\label{B-6}\structindent\it
  Let $A\in\H_{\text{loc}}(\rz^d)$, $V\in\K_{\pm}(\rz^d)$,
  $\psi\in\L^2(\rz^d)$, $\Lambda\subseteq\rz^d$ open, $t\ge 0$ and set for
  $n,m\in\nz$
   \begin{equation}\label{B.9}
      V_m^n(x) \defeq 
      \sup\{-n,\inf\{m,V(x)\}\}
   \end{equation}
   and consequently $V_m^\infty(x)=\inf\{m,V(x)\}$. Then
   \begin{equation}\label{B.10}
      \lim_{n\to\infty} 
      \left\|\,
         \left(
            \e^{-tH_\Lambda(A,V_n^m)} -
            \e^{-tH_\Lambda(A,V_\infty^m)} 
         \right)\psi
      \,\right\|_{2}
      = 0
   \end{equation}
   and
   \begin{equation}\label{B.11}
      \lim_{m\to\infty} 
      \left\|\,
         \left(
            \e^{-tH_\Lambda(A,V_\infty^m)}  -
            \e^{-tH_\Lambda(A,V)}
         \right)\psi
      \,\right\|_{2}
      = 0.
   \end{equation}
\end{structure}
\subsubsection*{Proof of Proposition~\ref{2-9}}
From \xref{B.10} and \xref{B.11} we conclude that for a suitably chosen
subsequence
\begin{equation}\label{B.12}
   \left(\e^{-t H_\Lambda(A,V_n^m)}\,\psi\right)(x)
   \to
   \left(\e^{-t H_\Lambda(A,V)}\,\psi\right)(x)
\end{equation}
as $n,m\to\infty$ for almost all $x\in\Lambda$. From Lemma~\ref{B-5} we know
that the Feynman-Kac-It\^o formula holds for the left-hand side of \xref{B.12}
and it therefore remains to show
\begin{equation}\label{B.13}\!\!\!\!\!\!\!\!\!\!\!\!\!\!\!\!
   \lim_{n,m\to\infty}\E_{x}\!\left[
       \e^{-S_t(A,V_n^m|w)}\,
       \Xi_{\Lambda,t}(w)\,\psi(w(t))
   \right]
   =
   \E_{x}\!\left[
       \e^{-S_t(A,V|w)}\,
       \Xi_{\Lambda,t}(w)\,\psi(w(t))
   \right]\!\!\!\!\!\!\!\!
\end{equation}
for almost all $x\in\Lambda$. To this end note that 
\begin{equation}\label{B.14}
   \left| \e^{-S_t(A,V_n^m|w)}\,\Xi_{\Lambda,t}(w) \right|
   \le \left| \e^{-S_t(0, V^-|w)} \right|
\end{equation}
and 
\begin{equation}\label{B.15}
   \E_{x}\!\left[
      \e^{-S_t(0,V^-|w)}
      \,| \psi(w(t))| 
   \right]
   \le
   \left\|\,
      \e^{-tH_{\rz^d}(0, V^-)} 
   \,\right\|_{2,\infty}
   < \infty
\end{equation}
by \xref{2.34} for $\Lambda=\rz^d$ and \Cite[Theorem~B.1.1]{Sim82}, see also
Lemma \ref{C-1}.  Employing the dominated-convergence theorem we conclude that
it is enough to establish
\begin{equation}\label{B.16}
   \lim_{n,m\to\infty} \int_0^t V_n^m(w(s)) \, \d s
   = \int_0^t V(w(s)) \, \d s
\end{equation}
for all $x=w(0)$ almost surely. Since $|V_n^m|\le|V|\in\K_{\pm}(\rz^d)$ this is
achieved by applying dominated convergence with the help of
Remark~\ref{2-8-ii}.\halmos
\subsubsection*{Proof of Lemma~\ref{B-1}}
Since $U_n^\Lambda$ and $V$ are bounded one has from \xref{2.23}
\begin{equation}\label{B.17}
   \Q\!\left(h_{\rz^d}^{A,\,\mu U_n^\Lambda+V}\right)
   = \Q\!\left(h_{\rz^d}^{A,0}\right)
   \supset \Q\!\left(h_{\rz^d}^{A,\,\mu U_\infty^\Lambda+V}\right)
   = \Q\!\left(h_{\rz^d}^{A,U_\infty^\Lambda}\right)
\end{equation}
and 
\begin{equation}\label{B.18}
   \Q\!\left(h_{\rz^d}^{A,U_\infty^\Lambda}\right)
   = 
   \left\{
      \psi\in \Q\!\left(h_{\rz^d}^{A,0}\right) :
      \sup_{n\in\nz} h_{\rz^d}^{A,U_n^\Lambda}(\psi,\psi) < \infty
   \right\}.
\end{equation}
Furthermore, we have by construction the monotonicity and boundedness
\begin{equation}\label{B.19}
   h_{\rz^d}^{A,\,\mu U_n^\Lambda+V} \le
   h_{\rz^d}^{A,\,\mu U_{n+1}^\Lambda+V} \le 
   h_{\rz^d}^{A,\,\mu U_\infty^\Lambda+V}, \quad n\in\nz,
\end{equation}
as well as the pointwise convergence
\begin{equation}\label{B.20}
   \lim_{n\to\infty} h_{\rz^d}^{A,\,\mu U_n^\Lambda+V}(\phi,\psi)
   = h_{\rz^d}^{A,\,\mu U_\infty^\Lambda+V}(\phi,\psi), \quad
   \phi,\psi\in\Q\!\left(h_{\rz^d}^{A,U_\infty^\Lambda}\right).
\end{equation}
These facts ensure that the monotone-convergence theorem for not necessarily
densely defined sesquilinear forms \Cite[Theorem~4.1]{Sim78b},
\Cite[Theorem~3.1.b)]{Wei84} is applicable and yields the generalized strong
resolvent convergence
\begin{equation}\label{B.21}\!\!\!\!\!\!\!\!
   \lim_{n\to\infty} 
   \left\|\,\left(
      \frac{1}{1+H_{\rz^d}(A,\mu U_n^\Lambda+V)} -
      \frac{1}{1+H_{\rz^d}(A,\mu U_\infty^\Lambda+V)}
   \right)\psi \,\right\|_{2}
   = 0
\end{equation}
for all $\psi\in\L^2(\rz^d)$.  This implies \Cite[Theorem~4.2]{Sim78b} the
strong convergence of $\left\{f\!\left(H_{\rz^d}(A,\mu
    U_n^\Lambda+V)\right)\right\}_{n\in\nz}$ for all functions $f$ of type
\xref{B.3}. Especially, \xref{B.5} follows.\halmos
\subsubsection*{Proof of Lemma~\ref{B-2}}
The Feynman-Kac-It\^o formula \xref{2.34} for $\Lambda=\rz^d$ tells
\begin{equation}\label{B.22}
  \left(\e^{-tH_{\rz^d}(A,\,\mu U_n^\Lambda+V)}\psi\right)(x) =
  \E_{x}\!\left[
      \e^{-S_t(A,\,\mu U_n^\Lambda+V|w)}\,\psi(w(t))
  \right]
\end{equation}
for almost all $x\in\rz^d$. From Lemma~\ref{B-1} we learn that by passing to a
suitable subsequence the left-hand side of \xref{B.22} tends to the left-hand
side of \xref{B.6} as $n\to\infty$ for almost all $x\in\rz^d$. For the proof of
the convergence of the right-hand side of \xref{B.22} to the desired limit we
only have to check
\begin{equation}\label{B.23}
   \e^{-S_t(0,\,\mu U_n^\Lambda|w)} \to
    \e^{-S_t(0,\,\mu U_\infty^\Lambda|w)} \,\Xi_{\Lambda,t}(w)
\end{equation}
as $n\to\infty$ for almost all $w$, because
\begin{equation}\label{B.24}
   \left| \e^{-S_t(A,\,\mu U_n^\Lambda +V|w)} \right| \le 1
\end{equation}
allows the application of the dominated-convergence theorem.

If $x=w(0)\not\in\overline\Lambda$ the left-hand side of \xref{B.23} tends to
$0$ for all continuous $w$. Therefore we may assume $x\in\overline\Lambda$. In
addition it is sufficient to consider a path $w$ which is H\"older continuous
of order $\frac{1}{3}$ \Cite[Theorem~5.2]{Sim79a}. Assume that $w$ leaves
$\Lambda$ and set $0\le\tau\le t$ to be the first exit time of $w$ from
$\Lambda$, that is
\begin{equation}\label{B.25}
   \tau \defeq \inf\{0< s\le t: w(s)\not\in\Lambda\} .
\end{equation}
Then by \xref{B.1}, \xref{B.2} 
\begin{equation}\label{B.26}
   \begin{split} \Ds
      S_t(0,\mu U_n^\Lambda|w) 
   \ge\; & \Ds
      \mu\int_0^t\inf\!\left\{|w(s)-w(\tau)|^{-3},n\right\} \d s
   \\ \Ds \ge\; & \Ds
      \mu\int_0^t\inf\!\left\{\frac{c}{|s-\tau|},n\right\} \d s
      \to\infty
   \end{split}
\end{equation}
as $n\to\infty$, where the constant $c>0$ depends on the path $w$. Thus
\xref{B.23} holds for almost all paths $w$ that do not stay in $\Lambda$. But
for a path $w$ staying in $\Lambda$ \xref{B.23} is a consequence of the
monotone-convergence theorem. Whence \xref{B.23} is true for almost all
$w$.\halmos
\subsubsection*{Proof of Lemma~\ref{B-3}}
By definition, $\Q\!\left(h_\Lambda^{A,0}\right)$ is the completion of
$\C_0^\infty(\Lambda)$ with respect to $\|\bullet\|_{h_{\rz^d}^{A,0}}$.
Moreover
$\C_0^\infty(\Lambda)\subset\Q\!\left(h_{\rz^d}^{A,U_\infty^\Lambda}\right)$.
Therefore, $\Q\!\left(h_\Lambda^{A,0}\right)$ is a subset of the completion of
$\Q\!\left(h_{\rz^d}^{A,U_\infty^\Lambda}\right)$.  To prove the converse
inclusion it is enough to establish that $\C_0^\infty(\Lambda)$ is dense in
$\Q\!\left(h_{\rz^d}^{A,U_\infty^\Lambda}\right)$ with respect to
$\|\bullet\|_{h_{\rz^d}^{A,0}}$. This in turn follows, confer the proof of
\Cite[Theorem~1.13]{CyFr87}, from the following three assertions:
\begin{itemize}
\item[{\bf 1)}] 
   $\L^\infty(\Lambda)\cap\Q\!\left(h_{\rz^d}^{A,U_\infty^\Lambda}\right)$
   is dense in
   $\Q\!\left(h_{\rz^d}^{A,U_\infty^\Lambda}\right)$.
\item[{\bf 2)}] 
   $\L^\infty_0(\Lambda)\cap
    \Q\!\left(h_{\rz^d}^{A,U_\infty^\Lambda}\right)$
   is dense in
   $\L^\infty(\Lambda)\cap\Q\!\left(h_{\rz^d}^{A,U_\infty^\Lambda}\right)$.
\item[{\bf 3)}] 
   $\C_0^\infty(\Lambda)$
   is dense in
   $\L^\infty_0(\Lambda)\cap
    \Q\!\left(h_{\rz^d}^{A,U_\infty^\Lambda}\right)$.
\end{itemize}
Here denseness is always meant with respect to the form norm
$\|\bullet\|_{h_{\rz^d}^{A,0}}$ and $\L^\infty_0(\Lambda)$ denotes the space of
bounded complex-valued functions with compact support inside $\Lambda$.
\subsubsection*{As to assertion 1)}
Fix $t>0$. Then $\e^{-t H_{\rz^d}(A,U_\infty^\Lambda)}\L^2(\Lambda)$ is an
operator core for $H_{\rz^d}(A,U_\infty^\Lambda)$ by the spectral theorem
\Cite[Section~VIII.8]{ReSi80}. Consequently it is also a form core for
$h_{\rz^d}^{A,U_\infty^\Lambda}$. From \xref{B.7} we conclude that
$\e^{-tH_{\rz^d}(A,U_\infty^\Lambda)}\L^2(\Lambda)\subseteq\L^\infty(\Lambda)$
and therefore
\begin{equation}\label{B.27}
   \L^\infty(\Lambda)\cap
   \e^{-tH_{\rz^d}(A,U_\infty^\Lambda)}\L^2(\Lambda)
   \subseteq
   \L^\infty(\Lambda)\cap
   \Q\!\left(h_{\rz^d}^{A,U_\infty^\Lambda}\right)
\end{equation}
is dense in $\Q\!\left(h_{\rz^d}^{A,U_\infty^\Lambda}\right)$ with
respect to $\|\bullet\|_{h_{\rz^d}^{A,U_\infty^\Lambda}}$. Since 
$\|\bullet\|_{h_{\rz^d}^{A,0}}\le\|\bullet\|_{h_{\rz^d}^{A,U_\infty^\Lambda}}$
the assertion follows.
\subsubsection*{As to assertion 2)}
Pick $\psi\in\L^\infty(\Lambda)\cap
\Q\!\left(h_{\rz^d}^{A,U_\infty^\Lambda}\right)$, then
$\theta_l\psi\in\L^\infty_0(\Lambda)\cap
\Q\!\left(h_{\rz^d}^{A,U_\infty^\Lambda}\right)$ for all $l\in\nz$. We want to
show $\|\psi-\theta_l\psi\|_{h_{\rz^d}^{A,0}}\to 0$ as $l\to\infty$. By the
construction of $\theta_l$ we know that $\|(1-\theta_l)\phi\|_{2}\to 0$ for all
$\phi\in\L^2(\Lambda)$.  Therefore, it remains to show
\begin{equation}\label{B.28}
   \lim_{l\to\infty} \|(-i\nabla-A)(\psi-\theta_l\psi)\|_{2} = 0.
\end{equation}
The triangle inequality gives
\begin{equation}\label{B.29}
   \|(-i\nabla-A)(\psi-\theta_l\psi)\|_{2} \le
   \|(1-\theta_l)(-i\nabla-A)\psi\|_{2} +
   \|\psi\nabla\theta_l\|_{2}.
\end{equation}
The first term on the right-hand side tends to $0$ as $l\to\infty$, because
$(-i\nabla-A)\psi\in\left(\L^2(\Lambda)\right)^d$ due to
$\psi\in\Q\!\left(h_{\rz^d}^{A,0}\right)$. We even know
$\psi\in\Q\!\left(h_{\rz^d}^{A,U_\infty^\Lambda}\right)$ and therefore
\begin{equation}\label{B.30}
   \sum_{l\in\nz} \|\psi\nabla\theta_l\|_{2}^2 \le
   h_{\rz^d}^{A,U_\infty^\Lambda}(\psi,\psi) <\infty
\end{equation}
which implies $\|\psi\nabla\theta_l\|_{2}\to 0$ as $l\to\infty$.
\subsubsection*{As to assertion 3)}
Pick $\psi\in\L^\infty_0(\Lambda)\cap
\Q\!\left(h_{\rz^d}^{A,U_\infty^\Lambda}\right)$, then
$\delta_r\ast\psi\in\C_0^\infty(\Lambda)$ for suitably small $r>0$, where
$\delta_r$ is the approximate $\delta$-function defined in Appendix~\ref{A}.
Since $\lim_{r\downarrow 0}\|\phi-\delta_r\ast\phi\|_{2}=0$ for all
$\phi\in\L^2(\Lambda)$ we aim to show
\begin{equation}\label{B.31}
   \|(-i\nabla-A)(\psi-\delta_r\ast\psi)\|_{2} \to 0
\end{equation}
as $r\downarrow 0$ at least for a suitably chosen subsequence.

We first claim that $\nabla\psi\in\left(\L^2(\Lambda)\right)^d$. This follows
from the fact that on the one hand $A\psi\in\left(\L^2(\Lambda)\right)^d$
because
$A\in\left(\L^2_{\text{loc}}(\Lambda)\right)^d\supset\H_{\text{loc}}(\rz^d)$
and $\psi\in\L^\infty_0(\Lambda)$ and on the other hand
$(-i\nabla-A)\psi\in\left(\L^2(\Lambda)\right)^d$ because
$\psi\in\Q\!\left(h_{\rz^d}^{A,0}\right)$. Therefore
\begin{equation}\label{B.32}
   \|(-i\nabla)(\psi-\delta_r\ast\psi)\|_{2} =
   \|\nabla\psi-\delta_r\ast(\nabla\psi)\|_{2} \to 0
\end{equation}
as $r\downarrow 0$.

Let $K\subset\Lambda$ denote the compact support of $\delta_R\ast|\psi|$ for
some small $R>0$. Then for all $0<r\le R$
\begin{equation}\label{B.33}
   |A(x)(\psi-\delta_r\ast\psi)(x)| \le
   2\|\psi\|_\infty\,\chi_{_{\scriptstyle K}}(x)\,|A(x)|.
\end{equation}
The right-hand side of this estimate -- considered as a function of $x$ -- is
in $\L^2(\Lambda)$ because $A\in\left(\L^2_{\text{loc}}(\Lambda)\right)^d$.
Moreover, we may choose a subsequence such that $(\psi-\delta_r\ast\psi)(x)\to
0$ for almost all $x\in\Lambda$ and then get from the dominated-convergence
theorem
\begin{equation}\label{B.34}
   \|A(\psi-\delta_r\ast\psi)\|_{2} \to 0
\end{equation}
for this subsequence.

The convergences \xref{B.32} and \xref{B.34} imply \xref{B.31} and hence
assertion~3.\halmos
\subsubsection*{Proof of Lemma~\ref{B-4}}
Recall that
\begin{equation}\label{B.35}\!\!\!\!\!\!\!\!\!\!\!\!\!\!\!\!
   \C_0^\infty(\Lambda)\subset
   \Q\!\left(h_{\rz^d}^{A,\,\mu U_\infty^\Lambda+V}\right) =
   \Q\!\left(h_{\rz^d}^{A,U_\infty^\Lambda}\right) \subseteq
   \Q\!\left(h_\Lambda^{A,0}\right) =
   \Q\!\left(h_\Lambda^{A,V}\right) \subset
   \L^2(\Lambda)
\end{equation}
independent of $\mu>0$, where the inclusion follows from Lemma~\ref{B-3}.
Therefore the monotonicity statement
\begin{equation}\label{B.36}
   h_{\rz^d}^{A,\,\mu'U_\infty^\Lambda+V}(\psi,\psi) \ge
   h_{\rz^d}^{A,\,\mu U_\infty^\Lambda+V}(\psi,\psi) \ge
   h_\Lambda^{A,V}(\psi,\psi) 
   , \quad \mu'>\mu,
\end{equation}
and the convergence
\begin{equation}\label{B.37}
   \lim_{\mu\downarrow 0}    
    h_{\rz^d}^{A,\,\mu U_\infty^\Lambda+V}(\psi,\psi) =
   h_\Lambda^{A,V}(\psi,\psi) 
\end{equation}
make sense for all $\psi\in\Q\!\left(h_{\rz^d}^{A,U_\infty^\Lambda}\right)$
and follow from the construction of the forms. We conclude with the help of
Lemma~\ref{B-3} that the monotone-convergence theorem for densely defined forms
\Cite[Theorem~3.2]{Sim78b}, \Cite[Theorem~S.16]{ReSi80}, see also
\Cite[Theorem~3.1.a)]{Wei84}, is applicable and yields the strong resolvent
convergence of $H_{\rz^d}(A,\mu U_\infty^\Lambda+V)$ to $H_\Lambda(A,V)$ as
$\mu\downarrow 0$. According to \Cite[Theorem~VIII.20]{ReSi80} this implies
\xref{B.8}.\halmos
\subsubsection*{Proof of Lemma~\ref{B-6}}
First we note that $V_n^m\in\L^\infty(\rz^d)$ and
$V_\infty^m\in\K(\rz^d)$ implies
\begin{equation}\label{B.38}
   \Q\!\left(h_\Lambda^{A,V_n^m}\right) =
   \Q\!\left(h_\Lambda^{A,0}\right) =
   \Q\!\left(h_\Lambda^{A,V_\infty^m}\right).
\end{equation}
Moreover, we know that $h_\Lambda^{A,V_\infty^m}$ is bounded from below.

By construction of $V_n^m$ we have the monotonicity
\begin{equation}\label{B.39}
   h_\Lambda^{A,V_n^m}(\psi,\psi) \ge
   h_\Lambda^{A,V_{n+1}^m}(\psi,\psi) \ge
   h_\Lambda^{A,V_\infty^m}(\psi,\psi) 
\end{equation}
and 
\begin{equation}\label{B.40}
   h_\Lambda^{A,V_\infty^m}(\psi,\psi)  =
   \lim_{n\to\infty}h_\Lambda^{A,V_n^m}(\psi,\psi) 
\end{equation}
for all $\psi\in\Q\!\left(h_\Lambda^{A,0}\right)$. These facts imply according
to \Cite[Theorem~3.2]{Sim78b}, \Cite[Theorem~S.16]{ReSi80} the strong resolvent
convergence of $H_\Lambda(A,V_n^m)$ to $H_\Lambda(A,V_\infty^m)$ as
$n\to\infty$.  By \Cite[Theorem~VIII.20]{ReSi80} Equation \xref{B.10} follows.

To prove \xref{B.11} we first note that 
\begin{equation}\label{B.41}
   \Upsilon \defeq 
   \left\{
      \psi\in   \Q\!\left(h_\Lambda^{A,0}\right) :
      \sup_{m\in\nz}h_\Lambda^{A,V_\infty^m}(\psi,\psi) < \infty 
   \right\}
\end{equation}
is dense in $\L^2(\Lambda)$ because $\C_0^\infty(\Lambda)\subset\Upsilon$.
Furthermore we have the monotonicity
\begin{equation}\label{B.42}
   h_\Lambda^{A,V_\infty^m}(\psi,\psi) \le
   h_\Lambda^{A,V_\infty^{m+1}}(\psi,\psi) \le
   h_\Lambda^{A,V}(\psi,\psi) 
\end{equation}
and 
\begin{equation}\label{B.43}
   h_\Lambda^{A,V}(\psi,\psi)  =
   \lim_{m\to\infty}h_\Lambda^{A,V_\infty^m}(\psi,\psi) 
\end{equation}
for all $\psi\in\Upsilon$. With the help of \Cite[Theorem~3.1]{Sim78b},
\Cite[Theorem~S.14]{ReSi80} we conclude that
$\Upsilon=\Q\!\left(h_\Lambda^{A,V}\right) =\Q\!\left(h_\Lambda^{A,V^+}\right)$
and the strong resolvent convergence of $H_\Lambda(A,V_\infty^m)$ to
$H_\Lambda(A,V)$ as $m\to\infty$. Thus \xref{B.11} follows by using
\Cite[Theorem~VIII.20]{ReSi80} again.\halmos
\Section{Brownian-motion estimates}
\label{C}
In this appendix we have gathered the probabilistic estimates which play a key
r\^ole in the proofs of Theorems~\ref{4-1}, \ref{4-3}, \ref{4-4}, \ref{5-1},
\ref{5-5}, \ref{6-1}, \ref{6-3} and \ref{6-5}. Our proofs of these estimates
are inspired by ideas for zero vector potentials in \Cite[Section~III]{Car79},
\Cite[\S~B.1,~\S~B.10]{Sim82} and the more recent monograph
\Cite[Section~3.2]{ChZh95}.
\begin{structure}{Lemma}\label{C-1}\structindent\it
   Let
   $\left\{V_n\right\}_{n\in\nz}\subset\K_{\pm}(\rz^d)$
   obey \xref{5.7}. Then
   \begin{equation}\label{C.1}
      \sup_{\tau_1 \le t \le \tau_2} \sup_{n\in\nz}
      \left\|\,
         \e^{-tH_{\rz^d}(0,V_n)}
      \,\right\|_{p,q}
      < \infty
   \end{equation}
   for all $0 < \tau_1 \le \tau_2 < \infty$ and $1 \le p \le q \le
   \infty$. Furthermore, for $p=q$ one may allow $\tau_1=0$.
\end{structure}
\subsubsection*{Proof:}
The Riesz-Thorin interpolation theorem \Cite[Theorem~IX.17]{ReSi75} and the
self-adjointness of the semigroup imply that it is enough to prove the Lemma
for the two cases $p=q=\infty$, $\tau_1=0$ and $p=1$, $q=\infty$, $\tau_1>0$.

In order to show this we first observe that
\begin{equation}\label{C.2}
   \left\|\,
      \e^{-2tH_{\rz^d}(0,V_n)}
   \,\right\|_{1,\infty}
   \le
   \left(
      \left\|\,
         \e^{-tH_{\rz^d}(0,V_n)}
      \,\right\|_{2,\infty}
   \right)^2
\end{equation}
holds by the semigroup property, the inequality \xref{5.18} and by
self-adjointness. Moreover, the inequality
\begin{equation}\label{C.3}
   \left\|\,
      \e^{-tH_{\rz^d}(0,V_n)}
   \,\right\|_{2,\infty}
   \le
   \left(
      2\pi t
   \right)^{-d/4}\:
   \left(
      \left\|\,
                 \e^{-tH_{\rz^d}(0,2V_n)}
      \,\right\|_{\infty,\infty}
   \right)^{1/2}
\end{equation}
follows from employing the Cauchy-Schwarz inequality in the Feynman-Kac-It\^o
formula and from the elementary estimate
\begin{equation}\label{C.4}
   \left|\,
      \left(
         \e^{-tH_{\rz^d}(0,0)} \, |\psi|^2
      \right)(x)\,
   \right|
   \le
   \left(
      2\pi t
   \right)^{-d/2}\,
   \left(\left\|
      \psi
   \right\|_2\right)^2.
\end{equation}
The combination of \xref{C.2} and \xref{C.3} shows that we are left to treat
the case $p=q=\infty$, $\tau_1=0$.

The semigroup property \xref{2.37} and the inequality \xref{5.18} yield
\begin{equation}\label{C.5}
   \left\|\,
      \e^{-tH_{\rz^d}(0,V_n)}
   \,\right\|_{\infty,\infty}
   \le
   \left(
      \left\|\,
         \e^{-tH_{\rz^d}(0,V_n)/N}
      \,\right\|_{\infty,\infty}
   \right)^N,\quad N\in\nz.
\end{equation}
Therefore it suffices to prove the case $p=q=\infty$, $\tau_1=0$ for some small
$\tau_2>0$. By \xref{5.7} and Remark~\ref{5-2-i} we may assume $\tau_2$ small
enough to ensure
\begin{equation}\label{C.6}
   \alpha\defeq
   \sup_{n\in\nz}\,\sup_{x\in\rz^d}
   \E_{x}\!\left[\,
      \int_0^{\tau_2}
         V_n^-(w(s))
      \d s
   \right]
   <1.
\end{equation}
Employing Khas'minskii's lemma \Cite{Kha59}, \Cite[Lemma~3.7]{ChZh95} we get
\begin{equation}\label{C.7}
   \E_{x}\!\left[\,
      \e^{-S_t(0,V_n|w)}\,
   \right]
   \le
   \frac 1{1-\alpha}
\end{equation}
for all $0\le t\le\tau_2$. By the Feynman-Kac-It\^o formula it follows that
\begin{equation}\label{C.8}
   \left\|\,
      \e^{-tH_{\rz^d}(0,V_n)}
   \,\right\|_{\infty,\infty}
   \le
   \frac 1{1-\alpha}
\end{equation}
for all $0\le t\le\tau_2$ and $n\in\nz$.\halmos
A glance at the preceding proof reveals that \xref{C.7} not only holds for
almost all $x\in\rz^d$ but for all $x\in\rz^d$. We state a frequently used
consequence of this separately, see, for example,
\Cite[Proposition~3.8]{ChZh95}.
\begin{structure}{Lemma}\label{C-2}\structindent\it
   Let $V\in\K_{\pm}(\rz^d)$. Then
   \begin{equation}\label{C.9}
      \sup_{0 \le t \le \tau} \sup_{x\in\rz^d}\;
      \E_x\!\left[\,
         \e^{-S_t(0,V|w)}
      \,\right]
      < \infty
   \end{equation}
   for all $\tau>0$.
\end{structure}
For our purposes it is essential to extend a previously known result for $A=0$
\cite[Proposition~3.9]{ChZh95} to $A\neq 0$.
\begin{structure}{Lemma}\label{C-3}\structindent\it
   Let $A\in\H(\rz^d)$ and $V\in\K(\rz^d)$.
   Then
   \begin{equation}\label{C.10}
      \lim_{t\downarrow 0} \sup_{x\in \rz^d}
      \E_{x}\!\left[\,
         \left|
            1-\e^{-S_t(A,V|w)}
         \right|^p\,
      \right]
      = 0
   \end{equation}
   for all $0<p<\infty$.
\end{structure}
\unskip
\begin{structure}{Lemma}\label{C-4}\structindent\it
   Let $A\in\H(\rz^d)$,
   $\left\{A_m\right\}_{m\in\nz}\subset\H(\rz^d)$, $V\in\K(\rz^d)$
   and $\left\{V_n\right\}_{n\in\nz}\subset\K(\rz^d)$ such that
   \begin{gather}\label{C.11}
          \lim_{m\to\infty}
          \left\|\,
                 \left(A-A_m\right)^2
          \,\right\|_{\K(\rz^d)}
          = 0,
   \\ \label{C.12}
          \lim_{m\to\infty}
          \left\|\,
         \nabla\cdot A-\nabla\cdot A_m
          \,\right\|_{\K(\rz^d)}
          = 0
   \end{gather}
   and
   \begin{equation}\label{C.13}
          \lim_{n\to\infty}
          \left\|\,
                V-V_n
          \,\right\|_{\K(\rz^d)}
          = 0.
   \end{equation}
   Then
   \begin{equation}\label{C.14}
      \lim_{m,n\to\infty} \sup_{0 \le t \le \tau} \sup_{x\in \rz^d}
      \E_{x}\!\left[\,
         \left|
            \e^{-S_t(A,V|w)}
            -\e^{-S_t(A_m,V_n|w)}
         \right|^p\,
      \right]
      = 0
   \end{equation}
   for all $0 \le \tau < \infty$ and $0<p<\infty$.
\end{structure}
\subsubsection*{Proof of Lemmas~\ref{C-3} and \ref{C-4}:}
We start from the inequality
\begin{equation}\label{C.15}
  \left|
     \e^z - \e^{z'}
  \right|
  \le
  2^{1+1/q} \: \left|  z - z' \right|^{1/q} \:
  \e^{\sup\left\{
     {\rm Re}z , {\rm Re}z'
  \right\}}                    
\end{equation}
valid for all $z,z'\in\cz$ and $1\le q\le\infty$. It can be inferred from the
elementary inequalities $1-2\,|u|^{2/q}\le\cos u\,$ and $\left| \e^u -1 \right|
\le |u|^{1/q}\,\e^{\sup\{u,0\}}$ for all $u\in\rz$.

Choosing $q$ such that $\gamma\defeq pq/(q-p) >0$ we get from \xref{C.15}
\begin{equation}\label{C.16}\!\!\!\!\!\!\!\!
   \begin{split} \Ds & \Ds
      2^{-2p}\:
      \E_{x}\!\left[\,
         \left|
            \e^{-S_t(A,V|w)}
            -\e^{-S_t(A_m,V_n|w)}
         \right|^p\,
      \right]
   \\ \Ds \le \; & \Ds
      \E_{x}\!\left[\,
         \left|
            S_t(A-A_m,V-V_n|w)
         \right|^{p/q}\:
         \e^{-pS_t(0,-V^- - V_n^-|w)}\,
      \right]
   \\ \Ds \le \; & \Ds
      \left( \E_{x}\!\left[\,
         \left|
            S_t(A-A_m,V-V_n|w)
         \right|\,
      \right] \right)^{p/q}\:
      \left( \E_{x}\!\left[\,
         \e^{-\gamma S_t(0,-V^- - V_n^-|w)}\,
      \right] \right)^{p/\gamma}
   \\ \Ds \le \; & \Ds
      \left( \E_{x}\!\left[\,
         \left|
            S_t(A-A_m,V-V_n|w)
         \right|\,
      \right] \right)^{p/q}
   \\ \Ds  & \Ds \hbox{} \times
      \left(
         \left\|\,
            \e^{-\tau H_{\rz^d}(0,-2\gamma V^-)}
         \,\right\|_{\infty,\infty}
         \:
         \sup_{n\in\nz}\,
         \left\|\,
               \e^{-\tau H_{\rz^d}(0,-2\gamma V^-_n)}
         \,\right\|_{\infty,\infty}
      \right)^{p/2\gamma}.
   \end{split}
\end{equation}
Here we have used the H\"older inequality for the second step and the
Cauchy-Schwarz inequality together with the Feynman-Kac-It\^o formula for the
third step. As remarked similarly in the proof of Theorem~\ref{5-5},
$V\in\K(\rz^d)$, $\{V_n\}_{n\in\nz}\subset\K(\rz^d)$ and \xref{C.13} imply
\xref{5.7}. Hence Lemma~\ref{C-1} can be applied to the right-hand side of
\xref{C.16}. This guarantees together with \xref{2.40} that it is sufficient to
show
\begin{equation}\label{C.17}
   \lim_{m,n\to\infty}\,\sup_{0\le t\le \tau}\,\sup_{x\in\rz^d}\,
   \E_{x}\!\left[\,
      \left|
         S_t(A-A_m,V-V_n|w)
      \right|\,
   \right]
   =0
\end{equation}
in order to prove Lemma~\ref{C-4}. Similarly, to verify Lemma~\ref{C-3} it is
enough to show
\begin{equation}\label{C.18}
   \lim_{t\downarrow 0}\,\sup_{x\in\rz^d}\,
   \E_{x}\!\left[\,
      \left|
         S_t(A,V|w)
      \right|\,
   \right]
   =0,
\end{equation}
as can be inferred from \xref{C.16} by putting there $A_m=0$ and $V_n=0$ and
appealing to \xref{2.40}.

To this end, the triangle inequality, the Jensen inequality and the isometry
for stochastic integrals \Cite[Theorem~4.2.5]{Fri75} or
\Cite[Proposition~3.2.10]{KaSh91} yield
\begin{equation}\label{C.19}
   \begin{split} \Ds & \Ds
      \E_{x}\!\left[\,
         \left|
            S_t(A-A_m,V-V_n|w)
         \right|\,
      \right]
   \\ \Ds \le \; & \Ds
      \left(
         \E_{x}\!\left[\,
            \int_0^t
               \left(A_m(w(s))-A(w(s))\right)^2
            \d s
         \right]
      \right)^{1/2}
   \\ \Ds & \Ds \hbox{} +
      \frac 12 \:
      \E_{x}\!\left[\,
         \int_0^t
            \left|
               (\nabla\cdot A_m)(w(s))-
               (\nabla\cdot A)(w(s))
            \right|
         \d s
      \right]
   \\ \Ds & \Ds \hbox{} +
      \E_{x}\!\left[\,
         \int_0^t
            \left|V_n(w(s))-V(w(s))\right|
         \d s
      \right].
   \end{split}
\end{equation}
Now \xref{C.17} is implied by further estimating the right-hand side of
\xref{C.19} with the help of
\begin{equation}\label{C.20}
   \E_{x}\!\left[\,
      \int_0^t
         f(w(s))
      \d s
   \right]
   \le
   \tau \,
   \E_{x}\!\left[\,
      \int_0^1
         f\left(\tau^{1/2}w(s)\right)
      \d s
   \right],
\end{equation}
valid for $f\ge 0$ and $0\le t \le \tau$, and \xref{2.29} in combination with
\xref{2.10}.

Eventually \xref{C.18} follows directly from \xref{C.19} for $A_m=0$ and
$V_n=0$ by observing \xref{2.28}.\halmos
\begin{structure}{Lemma}\label{C-5}\structindent\it
   Let $A\in\H_{\text{loc}}(\rz^d)$ and
   $V\in\K_{\pm}(\rz^d)$. Then
   \begin{equation}\label{C.21}
      \lim_{t\downarrow 0} \sup_{x\in K}
      \E_{x}\!\left[\,
         \left|
            1-\e^{-S_t(A,V|w)}
         \right|^p\,
      \right]
      = 0
   \end{equation}
   for all compact $K\subset\rz^d$ and $0<p<\infty$.
\end{structure}
\unskip
\begin{structure}{Lemma}\label{C-6}\structindent\it
   Under the hypotheses of Theorem~\ref{5-1} one has
   \begin{equation}\label{C.22}
      \lim_{m,n\to\infty} \sup_{0 \le t \le \tau} \sup_{x\in K}
      \E_{x}\!\left[\,
         \left|
            \e^{-S_t(A,V|w)}
            -\e^{-S_t(A_m,V_n|w)}
         \right|^p\,
      \right]
      = 0
   \end{equation}
   for all compact $K\subset\rz^d$, $0 \le \tau < \infty$ and
   $0<p<\infty$.
\end{structure}
\subsubsection*{Proof of Lemmas~\ref{C-5} and \ref{C-6}:}
By Lemma~\ref{C-1} and a standard localization technique we will reduce
Lemma~\ref{C-5} and Lemma~\ref{C-6} to Lemma~\ref{C-3} and Lemma~\ref{C-4},
respectively. For this purpose it is convenient to define $A_0\defeq 0$ and
$V_0\defeq 0$. Moreover, we introduce the abbreviation
\begin{equation}\label{C.23}
   \Upsilon_{p,R}(m,n,t,x)
   \defeq
   \E_{x}\!\left[\,
      \left|
         \e^{-S_t(A,V|w)}
         -\e^{-S_t(A_m,V_n|w)}
      \right|^p\,\Xi_{B_R,t}(w)
   \right].
\end{equation}

By the Cauchy-Schwarz inequality we obtain
\begin{equation}\label{C.24}
   \begin{split} \Ds 
      \Upsilon_{p,\infty}(m,n,t,x)
   \; \le \; & \Ds
      \left(
         \E_{x}\left[ 1-\Xi_{B_R,t}(w) \right]
      \right)^{1/2}\:
      \left(
         \Upsilon_{2p,\infty}(m,n,t,x)
      \right)^{1/2}
   \\ \Ds &\Ds \hbox{} +
         \Upsilon_{p,R}(m,n,t,x).
   \end{split}
\end{equation}
The use of $|z-z'|^{2p}\le 2^{2p}\left( |z|^{2p}+|z'|^{2p}\right)$ for
$z,z'\in\cz$ and of the Feynman-Kac-It\^o formula leads to
\begin{equation}\label{C.25}
\!\!\!\!\!\!\!\!\!\!\!\!\!\!\!\!\!\!\!\!\!\!\!\!\!\!\!\!\!\!\!\!
   \Upsilon_{2p,\infty}(m,n,t,x)
   \le
   2^{2p}\:
   \left(
      \left\|\,
         \e^{-\tau H_{\rz^d}(0,-2pV^-)}
      \,\right\|_{\infty,\infty}
      + \sup_{n\in\nz}\,
      \left\|\,
            \e^{-\tau H_{\rz^d}(0,-2pV^-_n)}
      \,\right\|_{\infty,\infty}
   \right). \!\!\!\!\!\!\!\!\!\!\!\!\!\!\!\!
\end{equation}
L\'evy's maximal inequality \Cite[Equation~(7.6$^\prime$)]{Sim79a} gives
\begin{equation}\label{C.26}
   \lim_{R\to\infty}\,\sup_{x\in K}
   \E_{x}\left[ 1-\Xi_{B_R,t}(w) \right]
   =0.
\end{equation}
Now we insert the estimate \xref{C.25} into \xref{C.24} and use Lemma~\ref{C-1}
and \xref{C.26} to observe that it is sufficient to show
\begin{equation}\label{C.27}
   \lim_{t\downarrow 0}\,\sup_{x\in\rz^d}\,
   \Upsilon_{p,R}(0,0,t,x) =0
\end{equation}
and
\begin{equation}\label{C.28}
   \lim_{m,n\to\infty}\,\sup_{0\le t\le \tau}\,\sup_{x\in\rz^d}\,
   \Upsilon_{p,R}(m,n,t,x) =0
\end{equation}
for all $R>0$ in order to prove Lemma~\ref{C-5} and~\ref{C-6}, respectively.

To this end, let $\Theta_{R}$ be as in Appendix~\ref{A}. Then \xref{A.2}
implies that $\Upsilon_{p,R}(m,n,t,x)$ as defined in \xref{C.23} does not
change its value when one replaces $A$, $A_m$, $V$ and $V_n$ by
$A\Theta_{R+1}$, $A_m\Theta_{R+1}$, $V\Theta_{R+1}$ and $V_n\Theta_{R+1}$,
respectively. This shows that \xref{C.27} follows from Lemma~\ref{C-3} and
\xref{C.28} from Lemma~\ref{C-4}, since $A\Theta_{R+1}\in\H(\rz^d)$,
$\left\{A_m\Theta_{R+1}\right\}_{m\in\nz}\subset\H(\rz^d)$,
$V\Theta_{R+1}\in\K(\rz^d)$ and
$\left\{V_n\Theta_{R+1}\right\}_{n\in\nz}\subset\H(\rz^d)$ by
assumption.\halmos
\begin{structure}{Lemma}\label{C-7}\structindent\it
   If $\Lambda\subseteq\rz^d$ is regular, then
   \begin{equation}\label{C.29}
     \lim_{r\downarrow 0}\, \sup_{|y-x|<r}
     \E_y\!\left[
       \Xi_{\Lambda,t}(w)
     \right] =0
   \end{equation}
   for all $x\in\partial\Lambda$, $t>0$. If $\Lambda$ is uniformly
   regular, then
   \begin{equation}\label{C.30}
     \lim_{r\downarrow 0}\, \sup_{x\in\partial\Lambda}\,
     \sup_{|y-x|<r}
     \E_y\!\left[
       \Xi_{\Lambda,t}(w)
     \right] =0
   \end{equation}
   for all $t>0$.
\end{structure}
\subsubsection*{Proof:}
The first assertion is not new. It is, for example, coded in
\Cite[Corollary~II.1.11]{Bas95}. Therefore, we only show the second assertion,
but remark that the appropriate simplification of the following proof works for
the first one, too.

We start from the simple estimate
\begin{equation}\label{C.31}
  \E_y\!\left[
    \Xi_{\Lambda,t}(w)
  \right] 
  \le
  \E_y\!\left[
    \Xi_{\Lambda,t-\tau}(w(\bullet+\tau))
  \right] 
\end{equation}
valid for all $0<\tau<t$. Using the Markov property of Brownian motion we get
\begin{equation}\label{C.32}
  \E_y\!\left[
    \Xi_{\Lambda,t-\tau}(w(\bullet+\tau))
  \right] 
  =
  (2\pi\tau)^{-d/2} \int \d z \, \e^{-(y-z)^2/2\tau}\,
  \E_z\!\left[
    \Xi_{\Lambda,t-\tau}(w)
  \right] .
\end{equation}
This shows that the right-hand side of \xref{C.31}, considered as a function of
$y\in\rz^d$, is uniformly continuous for all $0<\tau<t$.  Therefore,
\begin{equation}\label{C.33}
  \lim_{r\downarrow 0}\, \sup_{x\in\partial\Lambda}\,\sup_{|y-x|<r}
  \E_y\!\left[
    \Xi_{\Lambda,t}(w)
  \right] 
  \le 
  \sup_{x\in\partial\Lambda}
  \E_x\!\left[
    \Xi_{\Lambda,t-\tau}(w(\bullet+\tau))
  \right] 
\end{equation}
for all $0<\tau<t$. Now \xref{C.30} follows from the definition \xref{2.42} of
uniform regularity.\halmos
\begin{structure}{Lemma}\label{C-8}\structindent\it
  Let $A\in\H_{\text{loc}}(\rz^d)$. Then \xref{6.6} and \xref{6.7} hold true
  for all $t>0$, $x,y\in\rz^d$.
\end{structure}
\subsubsection*{Proof:}
For given $t>0$, $x,y\in\rz^d$ we choose $R>0$ such that $x,y\in B_R$. To show
\xref{6.6} we use the estimate
\begin{equation}\label{C.34}
   \begin{split} \Ds & \Ds
      \P_{0,x}^{t,y}\!\left\{
         \int_0^t
            \left(A(b(s))\right)^2
         \,\d s =\infty
      \right\}
  \\ \Ds \le \; &\Ds
      \P_{0,x}^{t,y}\!\left\{
         \int_0^t
            \left(
               \chi_{_{\scriptstyle B_R}}\!(b(s)) \,
               A(b(s))
            \right)^2
         \,\d s =\infty
      \right\}
      +
      \P_{0,x}^{t,y}\!\left\{
         \sup_{0\le s\le t } \, |b(s)|\ge R
      \right\} .
  \end{split}
\end{equation}
Since
\begin{equation}\label{C.35}
   \lim_{R\to\infty}
   \P_{0,x}^{t,y}\!\left\{
      \sup_{0\le s\le t } \, |b(s)|\ge R
   \right\} =0 ,
\end{equation}
it is enough to show that
\begin{equation}\label{C.36}
   \E_{0,x}^{t,y}\!\left[
      \int_0^t
         \left|f(b(s))\right|
      \,\d s 
   \right]
   <\infty
\end{equation}
for all $f\in\K(\rz^d)$. Using \xref{6.5} and the time-reversal symmetry of the
Brownian bridge, one has
\begin{equation}\label{C.37}\!\!\!\!\!\!\!\!
   \E_{0,x}^{t,y}\!\left[
      \int_0^t
         \left|f(b(s))\right|
      \,\d s 
   \right]
   \le
   2^{1+d/2}\,\e^{(x-y)^2/2t}
   \sup_{z\in\rz^d}
   \E_{z}\!\left[
      \int_0^{t/2}
         \left|f(w(s))\right|
      \,\d s 
   \right] .
\end{equation}
Due to \xref{2.28} the right-hand side is finite and \xref{6.6} is therefore
established.

By a reasoning analogous to the one leading to the condition \xref{C.36} it is
sufficient to check \xref{6.7} for $A\in\H(\rz^d)$ in order to prove it for
$A\in\H_{\text{loc}}(\rz^d)$. To this end, we observe
\begin{equation}\label{C.38}\!\!\!\!\!\!\!\!\!\!\!\!\!\!\!\!
   \begin{split} \Ds & \Ds 
      2^{-d/2}\,\e^{-(x-y)^2/2t}
      \left|
         \E_{0,x}^{t,y}\!\left[
            \int_0^t
               A(b(s))\cdot
               \frac{y-b(s)}{t-s}
            \,\d s
         \right]
      \right|
  \\ \Ds \le\;  &\Ds
      \left|
         \E_{x}\!\left[
            \int_0^{t/2}
               A(w(s))\cdot
               \frac{y-w(s)}{t-s}
            \,\d s
         \right]
      \right|
      +
      \left|
         \E_{y}\!\left[
            \int_0^{t/2}
               A(w(s))\cdot
               \frac{y-w(s)}{s}
            \,\d s
         \right]
      \right| .
  \end{split}\!\!\!\!\!\!\!\!
\end{equation}
Here we have again used \xref{6.5} and the time-reversal symmetry.  The first
expectation on the right-hand side is seen to be finite for $A^2\in\K(\rz^d)$
by the Cauchy-Schwarz inequality. The second expectation is finite for
$\nabla\cdot A\in\K(\rz^d)$ due to the partial-integration identity
\begin{equation}\label{C.39}
   \E_{y}\!\left[
      A(w(s))\cdot
      \frac{y-w(s)}{s}
   \right]
   =
   - \E_{y}\!\left[
      (\nabla\cdot A)(w(s))
   \right] .
\end{equation}
Hence we are done.\halmos
\end{appendix}
\section*{Acknowledgement}\label{NoPolishAcknowledgement}
The authors are grateful to Karl-Theodor Sturm (Bonn, Germany) for clarifying
discussions. D.H.\ is much indebted to Werner Kirsch (Bochum, Germany) for
constant encouragement and stimulating discussions.  Furthermore, K.B.\ and
D.H.\ thank the Deutsche Forschungsgemeinschaft for financial support under
grant-no.\ Br 1894/1-1 and Hu$\,$773/1-1, respectively.
\def\bibit#1#2{\bibitem{#2}} \def\platz{99}

\def\indexname{Citation index}\input{x.ind}
\end{document}